\documentclass[journal]{IEEEtran}

\usepackage{graphicx}
\usepackage{subcaption}
\usepackage{cite}
\usepackage{amsmath}
\usepackage{bm}
\usepackage{amsfonts}
\usepackage{array, booktabs}
\usepackage{url}
\usepackage{stfloats}
\usepackage{enumitem} 
\usepackage{xcolor,soul,framed}
\usepackage{mdwmath}
\usepackage{mdwtab}
\usepackage{lineno,hyperref}

\usepackage{amssymb}
\usepackage{graphicx}
\usepackage{algorithm}
\usepackage{algpseudocode}
\usepackage{multirow}
\usepackage{booktabs}
\usepackage{cancel}

\begin{document}

\title{Image Semantic Communication with Quadtree Partition-based Coding}

\author{Yinhuan Huang,~\IEEEmembership{Student Member,~IEEE,}
        Zhijin Qin,~\IEEEmembership{Senior Member,~IEEE}

\thanks{Yinhuan Huang and Zhijin Qin are with the Department of Electronic Engineering, Tsinghua University, Beijing 100084, China, and with the State Key Laboratory of Space Network and Communications, Beijing, 100084, China.
(email: huangyh24@mails.tsinghua.edu.cn; qinzhijin@tsinghua.edu.cn).}
\thanks{Our project will be available at \url{https://github.com/hyh-bingo/Quad-LIC_Quad-DeepSC}.}
}


\maketitle

\begin{abstract}
  Deep learning based semantic communication (DeepSC) system has emerged as a promising paradigm for efficient wireless transmission. However, existing image DeepSC methods, frequently encounter challenges in balancing rate-distortion performance and computational complexity, and often exhibit inferior performance compared to traditional schemes, especially on high-resolution datasets. 
  To address these limitations, we propose a novel image DeepSC system, using quadtree partition-based joint semantic-channel coding, named Quad-DeepSC, which maintains low complexity while achieving state-of-the-art transmission performance. Based on maturing learned image compression technologies, we establish a unified DeepSC system design and training pipeline. 
  The proposed Quad-DeepSC integrates quadtree partition-based entropy estimation and feature coding modules with lightweight feature extraction and reconstruction networks to form an end-to-end architecture. During training, all components except the feature coding modules are jointly optimized as a compact learned image codec, Quad-LIC, for source compression tasks. The pretrained Quad-LIC is then embedded into Quad-DeepSC and fine-tuned end-to-end over wireless channels. Extensive experimental results demonstrate that Quad-DeepSC is the first DeepSC system to surpass conventional communication systems, which employ VTM for source coding and adopt the optimal MCS index under 3GPP standards for channel coding and digital modulation, in performance across datasets of varying resolutions. 
  Notably, both Quad-DeepSC and Quad-LIC exhibit minimal latency, rendering them well-suited for deployment in real-time wireless communication systems.
\end{abstract}

\begin{IEEEkeywords}
Deep learning, semantic communication, image transmission, image compression.
\end{IEEEkeywords}

\section{Introduction}
\label{sec:introduction}

\IEEEPARstart{I}{n} recent years, the vision of the sixth-generation (6G) mobile communication system has emerged, aiming for a deep integration of artificial intelligence (AI) and communications, with an emphasis on ultra-reliable, low-latency transmission and extended functional capabilities beyond the fifth generation (5G)~\cite{10639525, Computing, wang2023road}. Meanwhile, the explosive growth of digital traffic and the rising demand for ultra-large-scale image and video interactions have imposed stringent requirements on existing communication systems~\cite{ling2022future}, necessitating efficient transmission schemes to lower the network traffic and mitigate channel impairments.

Conventional communication systems primarily emphasize source coding design in an effort to meet the escalating demands. Traditional lossy image compression standards, including better portable graphics (BPG), and versatile video coding (VVC), have achieved remarkable compression performance. Among these, VVC is widely regarded as one of the most advanced manually designed codecs in terms of reconstruction quality, as measured by peak signal-to-noise ratio (PSNR) and multi-scale structural similarity (MS-SSIM)~\cite{he2022elic}. However, this advancement comes at the expense of increased computational complexity. As the number of partitioning modes and prediction directions increases, the encoding latency for searching the optimal rate-distortion (R-D) configuration grows, and the potential for further optimization becomes limited~\cite{dcvc-dc}.

Data-driven AI technologies enable two main approaches for image source coding: one is to enhance traditional image codecs~\cite{zhao2020adaptive} and the other is to develop learned image codecs (LICs). The latter has undergone substantial development, as it employ neural networks to optimize the entire semantic coding process, including nonlinear transforms, latent distributions estimation, and quantization. LICs have demonstrated the ability to effectively capture spatial dependencies among image pixels. For instance, ELIC~\cite{he2022elic} outperforms VVC in still image compression and is regarded as a promising candidate for next-generation standards. However, the performance of existing LICs is still limited by the inherent trade-off between R-D performance and computational complexity. Additionally, source coding often optimize independently of transmission constraints. When LICs are adopted in conventional image communication systems, channel impairments and floating-point arithmetic inaccuracies across heterogeneous hardware~\cite{dcvc-rt} can severely degrade reconstruction quality, as LICs rely on probability estimates with extremely high precision for entropy coding, thereby limiting their applicability.

Deep learning based semantic communication (DeepSC) systems have rapidly emerged as a novel paradigm that fully incorporates AI and are capable of merging the physical layer of conventional communication systems~\cite{deepsc}. 
By adopting joint semantic-channel coding (JSCC), DeepSC systems can semantically perceive source information, extract task-relevant features, and apply importance-aware unequal error protection, thereby effectively addressing the limitations of conventional communication systems. 
The pioneering work~\cite{deepjscc} proposed the first image DeepSC system. However, it exhibited several limitations, such as fixed output symbol length per model, due to static bandwidth allocation. Building upon this framework, \cite{deepjscc-b} proposed a symbol-layering mechanism for progressive transmission and \cite{10065571} made the model bit-based, enabling integration into existing digital communication systems. In contrast, \cite{ntscc} adopted a different DeepSC framework by incorporating a LIC into the JSCC architecture for semantic feature extraction, entropy estimation, and reconstruction. 
\cite{ntscc-plus} enhanced this architecture by integrating a LIC with checkerboard partition-based entropy model~\cite{he2021checkerboard}, achieving superior performance that surpassed the image communication scheme, which employs VTM~\cite{vvc_vtm} (state-of-the-art engineered image codec of the VVC standard) for source coding and adopt the optimal modulation and scheme (MCS) index under the 3GPP standards~\cite{3gpp38214} for channel coding and digital modulation, on medium-resolution image datasets. \cite{mdjcm} introduced digital modulation into the framework proposed in~\cite{ntscc}, further boosting performance improvements. Similar to~\cite{ntscc}, \cite{plit} adopted a variational autoencoder-based LIC and outperformed~\cite{deepjscc-b} in progressive transmission. 

These developments of image DeepSC systems underscore a key insight: DeepSC system employs JSCC, which is built upon a LIC, significantly enhancing the transmission efficiency. However, not all LICs can be seamlessly integrated into a DeepSC framework. For wireless communication deployment, a DeepSC system must achieve high transmission quality and computational efficiency. Existing studies have not yet established effective methods to adapt LICs for DeepSC or to design and train DeepSC models when low-complexity, high-efficiency LICs are unavailable. The degree to which current LIC techniques can be leveraged remains an open issue.

To address these challenges, we revisit the entropy estimation, which fundamentally determines the efficiency of both LIC and its DeepSC extension. 
In entropy estimation, latent features are hierarchically partitioned into groups, where each group is conditioned on the estimation of previous ones to model the probability distribution of the current group. 
Over the years, various partitioning strategies have been explored to balance R-D performance and computational complexity. 
Traditional channel-wise autoregressive partitioning~\cite{minnen2018joint} achieves high accuracy but incurs heavy computation. 
Checkerboard partitioning~\cite{he2021checkerboard} simplifies spatial dependency into two sequential steps, improving speed but reducing accuracy. 
Hybrid methods~\cite{he2022elic} divide features into multiple channel groups with checkerboard refinement, balancing accuracy and speed. 
More recent approaches, such as quadtree partitioning~\cite{dcvc-dc,lu2022high} and quincunx partitioning~\cite{lin2023multistage}, further exploit joint spatial–channel correlations, achieving superior trade-offs.

Building upon these insights, we propose a unified design and training strategy for the DeepSC framework, emphasizing efficient entropy estimation and feature coding based on quadtree partitioning.
In this design, the feature coding module maps latent features to transmitted symbols following the same conditional and hierarchical structure used in entropy estimation.
This coherence ensures that symbol lengths align with their estimated probabilities, while inter-symbol dependencies enhance robustness against channel impairments.
The proposed modules are integrated with feature extraction and reconstruction networks to form an end-to-end DeepSC system, termed Quad-DeepSC.
From this architecture, a corresponding LIC, named Quad-LIC, is derived, consisting of feature extraction, entropy estimation, and reconstruction components.
Quad-LIC is first pretrained and subsequently embedded into Quad-DeepSC for end-to-end optimization over fading channels, achieving substantial gains in transmission efficiency. The main contributions of this work are summarized as follows:

\begin{enumerate}
  \item A unified design and training strategy is developed to bridge LIC and semantic communication. Rather than relying on pretrained black-box LICs, the method first constructs a DeepSC model based on efficient feature partitioning, from which specific modules are extracted to form a compact LIC. The LIC is trained on source compression tasks and subsequently embedded into DeepSC model for end-to-end transmission optimization, ensuring effective reuse of current LIC techniques and stable convergence under diverse channel conditions.

  \item The proposed Quad-DeepSC architecture incorporates a quadtree partition-based entropy estimation and feature coding mechanism, in which hierarchical spatial–channel grouping jointly drives entropy modeling and feature-to-symbol transformation. This end-to-end design enhances coding efficiency and supports adaptive semantic transformation guided by entropy and contextual information.

  \item An efficient LIC, Quad-LIC, is derived from the Quad-DeepSC architecture. It inherits the quadtree partition-based entropy estimation and achieves high compression performance with fast coding speed. The unified training pipeline allows Quad-LIC to operate both as a standalone codec and as an embedded module within Quad-DeepSC.

  \item Extensive experiments demonstrate that Quad-DeepSC achieves state-of-the-art transmission efficiency and fast coding speed compared with existing DeepSC systems, while Quad-LIC exhibits competitive compression performance and verifies the effectiveness of the proposed hierarchical design strategy.
\end{enumerate}

The rest of this paper is structured as follows. Section~\ref{sec:system_model} presents the system models of LIC and DeepSC. Section~\ref{sec:architecture} describes the architecture and implementations of Quad-LIC and Quad-DeepSC. Section~\ref{sec:experiments} presents the numerical experimental results. Finally, Section~\ref{sec:conclusion} concludes this paper.

\textit{Notation:} Scalars, vectors, and matrices are represented by lowercase (e.g., $x$), bold lowercase (e.g., $\mathbf{x}$), and bold uppercase letters (e.g., $\mathbf{X}$), respectively. Additionally, $\mathbb{C}^{m \times n}$ and $\mathbb{R}^{m \times n}$ represent the space of $m \times n$ complex and real matrices, respectively. $p_x$ is the probability density function for the continuous random variable $x$, and $P_{\bar{x}}$ is the probability mass function for the discrete random variable $\bar{x}$. $\mathbb{E}[\cdot]$ denotes the expectation operator, $||\cdot||_2$ denotes the euclidean norm and $\odot$ denotes the element-wise multiplication. $\theta$ denotes the neural network parameters, and subscripts indicate the corresponding module (e.g., $\theta_{g_a}$). $\mathcal{N}(x|\mu, \sigma^2)=(2\pi \sigma^2)^{-1/2} \exp\left(-\frac{(x-\mu)^2}{2\sigma^2}\right)$ represents a gaussian distribution, and $\mathcal{U}(y-u, y+u)$ denotes a uniform distribution centered at $y$ with range $[y-u, y+u]$.

\section{System Model and Problem Formulation}
\label{sec:system_model}
In this section, we presents the system models of LIC and DeepSC, including problem formulations, variable definitions, loss functions, and detailed descriptions of the processing pipelines.

\subsection{System Model of LIC}

\begin{figure*}[htbp]
  \centering
  \includegraphics[width=0.99\textwidth]{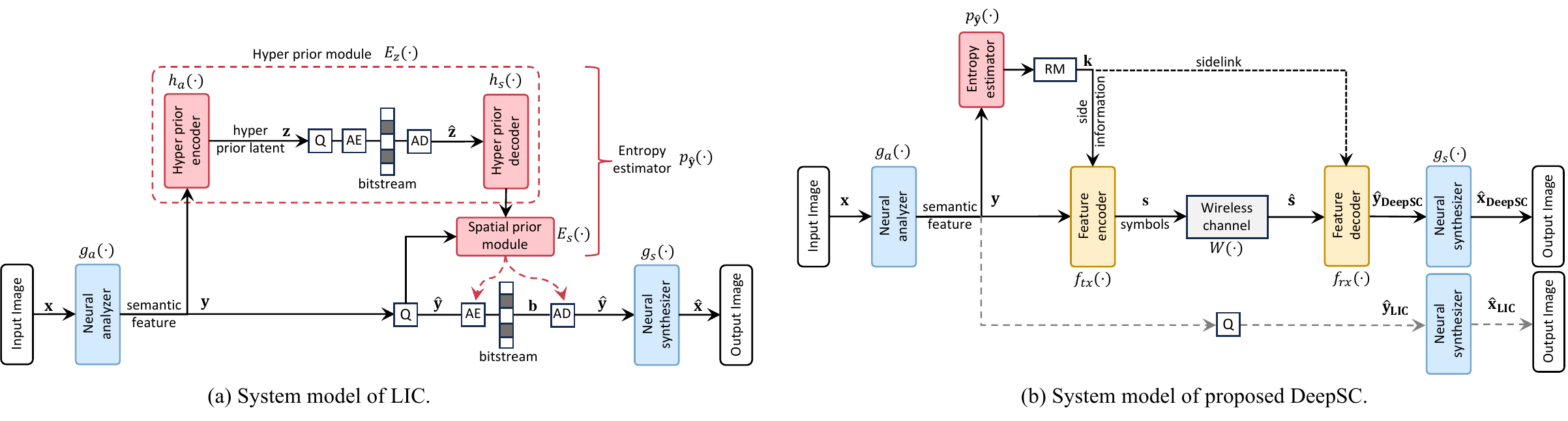}
  \caption{System models of LIC and DeepSC. (a) The architecture of LIC consists of a neural analyzer $g_a(\cdot)$, a neural synthesizer $g_s(\cdot)$, and an entropy estimator $p_{\mathbf{\hat{y}}}(\cdot)$. The entropy estimator includes a hyper prior module $E_z(\cdot)$ and a spatial prior module $E_s(\cdot)$. The hyper prior module contains an encoder $h_a(\cdot)$ and a decoder $h_s(\cdot)$. AE, AD, and Q represent arithmetic encoding, arithmetic decoding, and quantization, respectively. (b) The proposed DeepSC system extends the LIC framework by adding a feature encoder $f_{tx}(\cdot)$, a feature decoder $f_{rx}(\cdot)$, and a fixed, non-trainable channel model $W(\cdot)$. RM indicates the rate matching operation estimating symbol lengths. Procedures shown with dashed-gray lines in (b) are used solely to stabilize the training of DeepSC model. The identical naming indicates that the corresponding modules are the same.}
  \label{system_model}
\end{figure*}

LIC is a data-driven R-D optimization (RDO) framework. Consider an image vector $\mathbf{x} \in \mathbb{R}^{H \times W \times C}$, where $H$, $W$ and $C$ denote the height, width and number of color channels, respectively. As illustrated in Fig.~\ref{system_model}~(a), the LIC framework consists of a neural analyzer $g_a(\cdot)$, a neural synthesizer $g_s(\cdot)$, and an entropy estimator $p_{\mathbf{\hat{y}}}(\cdot)$. Given $\mathbf{x}$, the $g_a(\cdot)$ maps it to a semantic feature $\mathbf{y}$:
\begin{equation}
  \label{eq:g_a}
  \mathbf{y} = g_a(\mathbf{x}; \theta_{g_a}).
\end{equation}
The $\mathbf{y}$ is then quantized to $\mathbf{\hat{y}}=Q(\mathbf{y})$. Since $\mathbf{\hat{y}}$ is discrete, it can be compressed into a bitstream $\mathbf{b}$ using entropy coding techniques, such as arithmetic coding, and can be losslessly recovered from $\mathbf{b}$:
\begin{equation}
  \label{eq:AE}
    \mathbf{b} = AE(\mathbf{\hat{y}}, p_{\mathbf{\hat{y}}}(\mathbf{\hat{y}}; \theta_{p_{\mathbf{\hat{y}}}})),
\end{equation}
\begin{equation}
  \label{eq:AD}
    \mathbf{\hat{y}} = AD(\mathbf{b}, p_{\mathbf{\hat{y}}}(\mathbf{\hat{y}}; \theta_{p_{\mathbf{\hat{y}}}})),
\end{equation}
where $AE(\cdot)$ and $AD(\cdot)$ represent arithmetic encoding and arithmetic decoding. The $g_s(\cdot)$ is used to reconstruct the image $\mathbf{\hat{x}}$ from $\mathbf{\hat{y}}$:
\begin{equation}
  \label{eq:g_s}
  \mathbf{\hat{x}} = g_s(\mathbf{\hat{y}}; \theta_{g_s}).
\end{equation}

Due to the quantization process, reconstruction errors between $\mathbf{x}$ and $\mathbf{\hat{x}}$ are inevitable, which leads to the RDO problem. The loss function is expressed as:
\begin{equation}
  \label{eq:LIC_loss}
  L_{LIC}=\lambda \cdot \mathcal{D}\left(\mathbf{x}, \mathbf{\hat{x}} \right) + R,
\end{equation}
where $\lambda$ is a weighting coefficient, the distortion $\mathcal{D}(\cdot)$ is defined according to the downstream task, such as mean squared error (MSE) or MS-SSIM for image reconstruction, and the rate $R$ represents the length of $\mathbf{b}$, defined by cross entropy: 
\begin{equation}
  \label{eq:rate}
  R=\mathbb{E}_{\mathbf{x} \sim p_{\mathbf{x}}}\left[-\log _{2} P_{{\mathbf{\hat{y}}}}\left(Q\left(g_{a}\left(\mathbf{x}\right)\right)\right)\right].
\end{equation}

For training model, \cite{balle2016end} proposed using additive uniform noise to replace quantization:
\begin{equation}
  \label{eq:addnoise}
  \mathbf{\tilde{y}} = \mathbf{y} + \mathbf{u},
\end{equation}
where $\mathbf{u}$ denotes independent identically distributed (i.i.d.) samples from the $\mathcal{U}(-\frac{1}{2}, \frac{1}{2})$ for random quantization offsets. It is a technique widely adopted to address the zero gradient problem caused by quantization~\cite{pmlr-v139-guo21c, Pan_2021_CVPR}, though quantization is still employed in testing model. The RDO problem can be formally described as a variational autoencoder~\cite{balle2018variational}. $q(\mathbf{\tilde{y}}|\mathbf{x})$ estimated by $p_{\mathbf{\hat{y}}}(\cdot)$ is used to approximate the true posterior $p_{\mathbf{\tilde{y}}|\mathbf{x}}(\mathbf{\tilde{y}}|\mathbf{x})$ by minimizing the Kullback-Leibler (KL) divergence over the data distribution $p_{\mathbf{x}}(\cdot)$:
\begin{equation}
  \label{eq:map}
  \begin{aligned}
    \mathbb{E}_{\mathbf{x} \sim p_{\mathbf{x}}}\left[\mathcal{D}_{\mathrm{KL}}\left(q(\mathbf{\tilde{y}}|\mathbf{x}) \| p_{\mathbf{\tilde{y}}|\mathbf{x}}(\mathbf{\tilde{y}}|\mathbf{x})\right)\right] = \mathbb{E}_{\mathbf{x} \sim p_\mathbf{x}} \, \mathbb{E}_{\mathbf{\tilde{y}} \sim q(\mathbf{\tilde{y}}|\mathbf{x})} \\ 
    \Big[
      \cancelto{c_1}{\log q(\mathbf{\tilde{y}} \mid \mathbf{x})} 
      - \underbrace{\log p_{\mathbf{x} \mid \mathbf{\tilde{y}}}(\mathbf{x} \mid \mathbf{\tilde{y}})}_{\text{weighted distortion}} 
      - \underbrace{\log p_{\mathbf{\tilde{y}}}(\mathbf{\tilde{y}})}_{\text{rate}} 
      \Big] + c_2,
  \end{aligned}
\end{equation}
where $c_1$ and $c_2$ represent constants. Since the additive uniform noise aligns the center of the $\mathbf{\tilde{y}}$ with $\mathbf{y}$, in (\ref{eq:map}), the first term is a constant and can be dropped, as can the last term. The second term quantifies the distortion between $\mathbf{x}$ and $\mathbf{\hat{x}}$, which can be replaced by MSE. The third term corresponds to $R$ in (\ref{eq:rate}).

The $p_{\mathbf{\hat{y}}}(\cdot)$ comprises a hyper prior module $E_{z}(\cdot)$ and a spatial prior module $E_s(\cdot)$. The $E_z(\cdot)$ consists of a hyper prior encoder $h_a(\cdot)$ and a hyper prior decoder $h_s(\cdot)$. The $h_a(\cdot)$ learns the hyper prior latent $\mathbf{z}$, which is then quantized to $\mathbf{\hat{z}}=Q(\mathbf{z})$ and entropy coded with a learned factorized prior. For this, the $R$ in the loss function (\ref{eq:LIC_loss}) during training is updated as:
\begin{equation}
  \label{eq:rate2}
    R=R_y+R_z=\mathbb{E}_{\mathbf{x} \sim p_{\mathbf{x}}}\left[-\log _{2}p_{\mathbf{\tilde{y}} \mid  \mathbf{\tilde{z}}}(\mathbf{\tilde{y}} \mid  \mathbf{\tilde{z}}) -\log _{2}p_{\mathbf{\tilde{z}}}(\mathbf{\tilde{z}})\right].
\end{equation}
At the decoder side, we first use the $h_s(\cdot)$ to obtain the initial estimate parameters, which assists the $E_s(\cdot)$ in capturing the spatial dependencies. Under the joint influence of the $\mathbf{\hat{z}}$ and $g_a(\cdot)$, each element $y_i$ in $\mathbf{y}$ is modeled as an independent gaussian random variable, thus the conditional probability of $\mathbf{\tilde{y}}$ can be expressed as:
\begin{equation}
  \label{eq:prob_bpp_1}
    p_{\mathbf{\tilde{y}}}(\mathbf{\tilde{y}} \mid \mathbf{\hat{z}}) = \prod_i \left( \mathcal{N}(\mu_{i}, \sigma_{i}^2) * \mathcal{U}\left(-\frac{1}{2}, \frac{1}{2}\right) \right)(\tilde{y}_{i}),
\end{equation}
where
\begin{equation}
  \label{eq:prob_bpp_2}
  \mu_{i}, \sigma_{i} = E_s\left(h_s\left(\mathbf{\hat{z}}; \theta_{h_s}\right), \tilde{y}_{<i}; \theta_{E_s}\right).
\end{equation}

\subsection{System Model of DeepSC}
As for end-to-end wireless image transmission scenario, the DeepSC aims for low-latency transmission using minimal symbol sequences under channel constraints, facilitating efficient execution of downstream tasks~\cite{Task-Oriented}. Its architecture partially aligns with that of LIC. At the transmitter side, the neural analyzer $g_a(\cdot)$ extracts semantic features $\mathbf{y}$ from the source image $\mathbf{x}$. These features are subsequently mapped to complex-valued channel symbols $\mathbf{s} \in \mathbb{C}^{l}$ via a feature encoder $f_{tx}(\cdot)$:
\begin{equation}
  \label{eq:f_tx}
  \mathbf{s} = f_{tx}\Big( g_a(\mathbf{x}; \theta_{g_a}), p_{\mathbf{\hat{y}}}(g_a(\mathbf{x}; \theta_{g_a});\theta_{p_{\mathbf{\hat{y}}}}); \theta_{f_{tx}} \Big).
\end{equation}
At the receiver side, the feature decoder $f_{rx}(\cdot)$ reconstructs semantic features $\mathbf{\hat{y}_{DeepSC}}$ from the received symbols $\mathbf{\hat{s}}$, which are then fed into the synthesizer $g_s(\cdot)$ to recover the reconstructed image $\mathbf{\hat{x}_{DeepSC}}$. The proposed DeepSC system is illustrated in Fig.~\ref{system_model} (b). The dashed-gray-line procedures are only utilized during training to stabilize convergence. In this phase, arithmetic coding is omitted, and the quantization $Q(\cdot)$ is approximated by additive uniform noise to ensure differentiability.

The wireless channel is modeled by a transfer function $W(\mathbf{s})=\mathbf{h} \odot \mathbf{s} + \mathbf{n}$, where $\mathbf{h}\in \mathbb{C}^{l}$ is the complex channel coefficient sequence and $\mathbf{n}\!\sim\!\mathcal{CN}(\mathbf{0},\sigma^2\mathbf{I}_l)$ is additive white gaussian noise (AWGN). The signal-to-noise ratio (SNR) is defined as:
\begin{equation}
  \label{eq:snr}
  \begin{aligned}
     \text{SNR} = 10 \log_{10} \left( \frac{\mathbb{E}\left[\lVert \mathbf{h}\rVert_2^2\right]P}{\sigma^2} \right) 
      \geq \\
      10 \log_{10} \left( \frac{\mathbb{E}\left[\lVert \mathbf{h}\rVert_2^2\right]\mathbb{E}_{\mathbf{s}} \left[||\mathbf{s}||_2^2\right]}{\mathbb{E}_{\mathbf{n}} \left[||\mathbf{n}||_2^2\right]} \right),
  \end{aligned}
\end{equation}
where the $\sigma^2$ denotes the average noise power and $P$ represents the power constraint on the transmitted symbols $\mathbf{s}$, i.e., $\mathbb{E}_{\mathbf{s}} \left[||\mathbf{s}||_2^2\right] \leq P$. 
When the channel state information (CSI) $\mathbf{\hat{h}}$ is available, the receiver applies zero-forcing (ZF) equalization:
\begin{equation}
  \label{eq:zfimp}
  \hat{\mathbf{s}} \leftarrow \frac{\hat{\mathbf{h}}^*}{|\hat{\mathbf{h}}|^2} \odot W(\mathbf{s}),
\end{equation}
where the $\mathbf{\hat{h}}^*$ is the complex conjugate of $\mathbf{\hat{h}}$. In practice, CSI is imperfect. We adopt an additive CSI error model $\mathbf{\hat{h}} = \mathbf{h} + \mathbf{e}$, where $\mathbf{e}\sim\mathcal{CN}\!\left(0,\sigma_e^2\mathbf{I}_l\right)$. The estimation quality is measured by the normalized mean-squared error (NMSE)~\cite{bjornson2017massive}:
\begin{equation}
  \label{nmse}
  \mathrm{NMSE}_{\mathrm{dB}}=10\log_{10}(\frac{\mathbb{E}\left[\lVert \mathbf{e}\rVert_2^2\right]}{\mathbb{E}\left[\lVert \mathbf{h}\rVert_2^2\right]}).
\end{equation}

\begin{figure}[!t]
  \centering
  \includegraphics[width=3.3in]{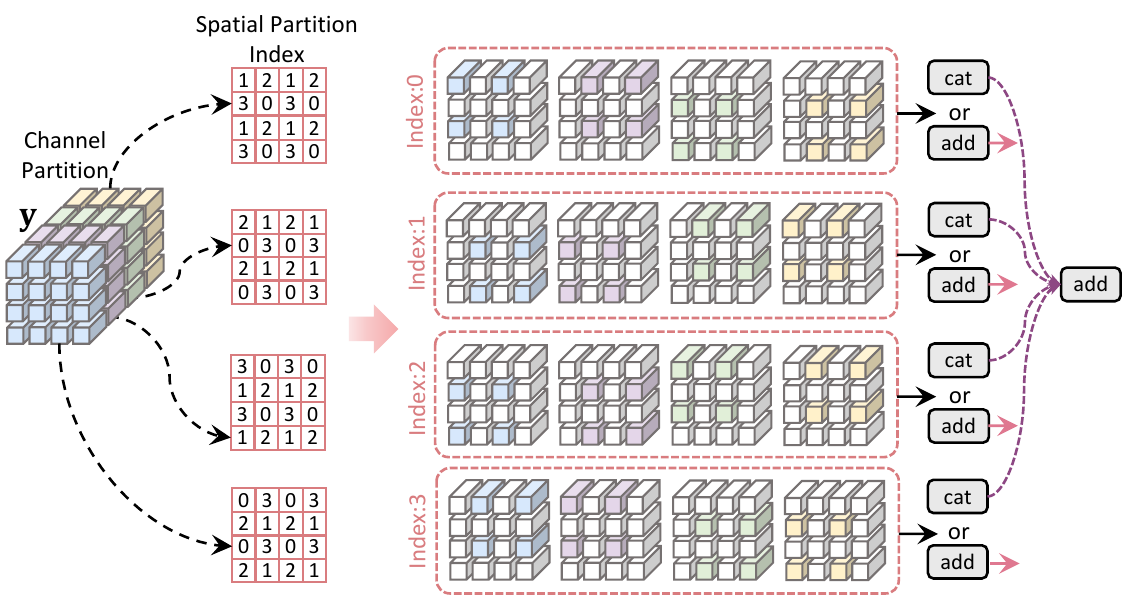}
  \caption{The detailed procedure of the quadtree partition, the inverse process is called quadtree fusion.}
  \label{Quadtree partition}
\end{figure}

We consider AWGN and fading channel scenarios. In the AWGN case, the channel coefficient $\mathbf{h}$ remains constant. For rayleigh fading, it is assumed that $\mathbf{h}$ varies independently for each transmission. In particular, $h_k\overset{\text{i.i.d.}}{\sim}\mathcal{CN}(0,1)$ for the $k$-th channel use. For correlated fading, $\mathbf{h}$ is generated through a first-order gauss–markov process: $h_k=\rho\,h_{k-1}+\alpha w_k$, where $w_k\sim\mathcal{CN}(0,1)$, $\alpha=\sqrt{1-\rho^2}$ and $\rho\in[0,1)$. The coefficient $\rho$ controls the correlation level ($\mathbb{E}[h_k h_{k+\Delta}^*]=\rho^{|\Delta|}$)~\cite{corrfading}. To capture burst errors, we let the noise variance follow a two-state Gilbert–Elliott markov process with transition probabilities $p$ (good$\to$bad) and $q$ (bad$\to$good), and in the bad state we additionally erase each symbol with probability $\varepsilon$. The instantaneous variance switches between $\sigma^2$ (good) and $\kappa\sigma^2$ (bad)~\cite{burst_error}. Random deep fading due to intermittent blockages is modeled by a markov switch on the channel itself: line-of-sight (LOS) segments use a rician model
$h_k=\mu+\sigma_s w_k$, where $w_k\sim\mathcal{CN}(0,1)$, $\mu=\sqrt{\tfrac{K}{K+1}}$ and $\sigma_s=\sqrt{\tfrac{1}{K+1}}$, while non-line-of-sight (NLOS) segments are rayleigh but attenuated in amplitude by $10^{-A_{\mathrm{blk}}^{(\mathrm{dB})}/20}$ and multiplied by a log-normal shadowing factor $10^{X_k/20}$, $X_k\sim\mathcal{N}(0,\sigma_{\mathrm{sh}}^2)$. Segment durations are drawn from geometric distributions to produce piecewise-constant runs along time~\cite{mmwaveblock0, mmwaveblock1}.

Following~\cite{deepjscc}, we denote the source bandwidth by $n=HWC$ and the channel bandwidth by $l$. The channel bandwidth ratio (CBR) is defined as $\text{CBR}\triangleq\frac{l}{n}=\frac{l}{HWC}$, measuring bandwidth efficiency as the number of channel uses per source pixel. The optimization of proposed DeepSC system is formulated as a RDO problem~\cite{ntscc}, analogous to (\ref{eq:LIC_loss}), and can be expressed as:
\begin{equation}
  \label{eq:DeepSC_loss}
  L_{DeepSC} = \lambda \cdot [\mathcal{D}\left(\mathbf{x}, \mathbf{\hat{x}_{LIC}} \right), + \mathcal{D}\left(\mathbf{x}, \mathbf{\hat{x}_{DeepSC}} \right)] + K,
\end{equation}
where $\lambda$ is a weighting coefficient, $\mathcal{D}(\cdot)$ denotes the distortion metric, and $K$ represents the rate term. The first distortion term, corresponding to the reconstructed image $\mathbf{\hat{x}_{LIC}}$ from the LIC, serves to stabilize the training process, shown in the dashed-gray line in Fig.~\ref{system_model} (b). The second term measures the distortion of the DeepSC output. The rate term $K$ quantifies the bandwidth cost and is estimated via LIC-based entropy modeling:

\begin{equation}
  \label{eq:K}
  \begin{aligned}
    K = \eta \cdot \Big( \mathbb{E}_{\mathbf{x} \sim p_{\mathbf{x}}} \mathbb{E}_{\mathbf{z} \sim p_{\mathbf{z}}}
      \Big[ - \log _{2} p_{\mathbf{y}|\mathbf{\tilde{z}}} (\mathbf{y}|\mathbf{\tilde{z}}) \Big]
      + \mathbb{E}_{\mathbf{k}} \left[b({\mathbf{k}})\right] \\
      + \underbrace{\mathbb{E}_{\mathbf{x} \sim p_{\mathbf{x}}} \mathbb{E}_{\mathbf{z} \sim p_{\mathbf{z}}}
      \Big[ -\log _{2} p_{\mathbf{\tilde{z}}} (\mathbf{\tilde{z}}) \Big]}_{\text{not actually transmitted}} \Big), 
  \end{aligned}
\end{equation}
where $\eta$ is the scaling hyperparameter that controls the spectral efficiency of transmission and $b(\mathbf{k})$ denotes the number of bits of the side information $\mathbf{k}$ after lossless compression. The $\mathbf{k}$ assists $f_{rx}$ in reconstructing the dimension corresponding to each symbol:
\begin{equation}
  \label{eq:f_rx}
  \mathbf{\hat{x}_{DeepSC}} = g_s\Big( f_{rx}(\mathbf{\hat{s}}, \mathbf{k}; \theta_{f_{rx}}); \theta_{g_s} \Big).
\end{equation}
In (\ref{eq:K}), the second item is applied only during the testing phase, as it does not contribute gradients in training, while the final term is included solely to stabilize training and is not transmitted.

The semantic feature $\mathbf{y} \in \mathbb{R}^{H_y \times W_y \times C_y}$ consists of $H_y \times W_y$ spatial units $\mathbf{y}_{sp}^{(i,j)}$, each represented by a $C_y$-dimensional vector and corresponds to a symbol $\mathbf{s}_{i,j} \in \mathbb{C}^{\lceil k_{i,j} / 2 \rceil}$, where the symbol length factor $k_{i,j}$ is computed as the entropy accumulation of $\mathbf{y}_{sp}^{(i,j)}$:
\begin{equation}
  \label{eq:kij}
  k_{i,j} = Q(\sum_{k=1}^{C_y} -\eta \log_{2} p_{y_{i,j,k}|\mathbf{\tilde{z}}} (y_{i,j,k}|\mathbf{\tilde{z}})).
\end{equation}
In practice, $k_{i,j}$ is quantized to the nearest value within a predefined discrete range and contributes to the overall side information vector $\mathbf{k} \in \mathbb{R}^{H_y \times W_y}$. This process is referred to as rate matching. The selection of this range balances the trade-off among search complexity, side information rate $\eta \cdot b(\mathbf{k})$, and reconstruction quality.

\begin{figure*}[htbp]
  \centering
  \includegraphics[width=0.99\textwidth]{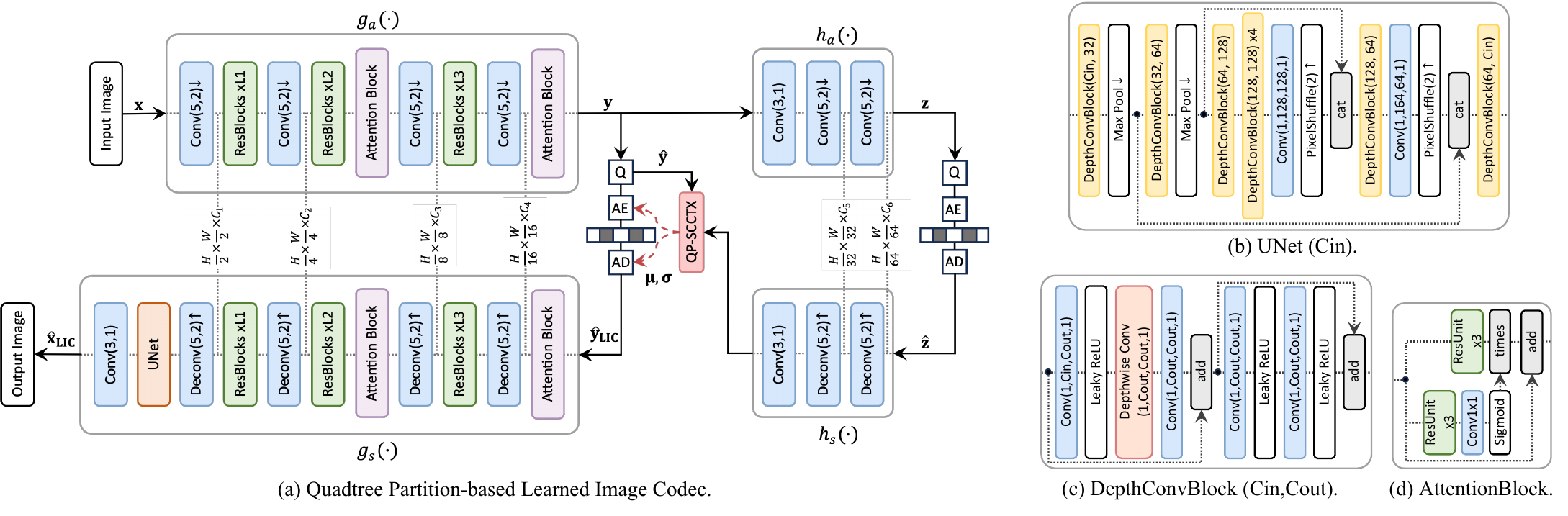}
  \caption{(a) Overview of proposed quadtree partition-based learned image codec (Quad-LIC). AE, AD, and Q denote the arithmetic encoding, arithmetic decoding, and quantization, respectively. QP-SCCTX is the proposed quadtree partition-based space-channel context model, illustrated in Fig.~\ref{entropy model}. (b) The network structure of the UNet. (c) The details of DepthConvBlock. (d) The details of AttentionBlock.}
  \label{overview of QSC}
\end{figure*}

\section{Architecture and Implementations}
\label{sec:architecture}
This section presents the architecture and implementation details of the proposed Quad-DeepSC. First, the integrated LIC-DeepSC strategy, which forms the foundation of the design, is introduced. Then, the model architecture of Quad-LIC and Quad-DeepSC are described. The quadtree partition-based coding scheme, which improves image transmission efficiency, is then detailed. Subsequently, the variable symbol length mapping technique and the transmission of associated side information are discussed. Finally, the training strategy used to optimize the entire framework is outlined.

\subsection{Integrated LIC-DeepSC Strategy}
The DeepSC system, built on LIC, can achieve excellent transmission performance. As illustrated in Fig.~\ref{system_model}, the neural analyzer $g_a(\cdot)$ and neural synthesizer $g_s(\cdot)$ in LIC learn an effective transformation between images and semantic features. The entropy estimator $p_{\mathbf{\hat{y}}}(\cdot)$ provides accurate probabilistic modeling, enabling efficient source compression. Based on the efficient LIC framework, DeepSC introduces a feature encoder $f_{tx}(\cdot)$ and a feature decoder $f_{rx}(\cdot)$ to learn a robust mapping between features and transmitted symbols, while leveraging $p_{\mathbf{\hat{y}}}(\cdot)$ for reliable symbol-length estimation. This joint design allows efficient conversion between images and variable-length symbol sequences.

However, this does not imply that all LIC are universally applicable. Although some LICs exhibited efficient source compression, their computational complexity restrict their applicability in real-time image transmission. Furthermore, when designing the DeepSC system based on LIC, both the $f_{tx}(\cdot)$ and $f_{rx}(\cdot)$ need to align with the $p_{\mathbf{\hat{y}}}(\cdot)$. Specifically, the reference content at each step of $p_{\mathbf{\hat{y}}}(\cdot)$ needs to correspond to that in $f_{tx}(\cdot)$ and $f_{rx}(\cdot)$. This underscores that LICs based on autoregressive entropy models involve numerous steps within their entropy estimation, making them suboptimal for this specific application. A mismatch between the symbol-length estimation and symbol-mapping processes can lead to performance degradation, as confirmed in Section~\ref{subsec:ablation studies}.

To enhance the design of the DeepSC system by leveraging advancements in LIC, we propose an integrated strategy. First, select a feature partitioning methods. In this paper, we adopt the quadtree partition~\cite{lu2022high, dcvc-dc}, as illustrated in Fig.~\ref{Quadtree partition}. This partitioning method divides the feature $\mathbf{y}$ into four parts along the channel dimension, with each part segmented step-by-step according to its corresponding spatial partition index. For instance, in step 0, we extract elements with index 0 from each part and then combine them through concatenation or addition to form the partitioned feature $\mathbf{y}_0$, where concatenation is employed for entropy estimation, as it requires maintaining spatially-related positions, whereas addition is used for feature coding. The advantage of this approach lies in its reduced step count and its ability to effectively guide $f_{tx}(\cdot)$ and $p_{\mathbf{\hat{y}}}(\cdot)$ in removing spatial correlation from each element of $\mathbf{y}$ while maintaining parallelism. Next, we design the feature codec and entropy model, where the input at each step $i$ corresponds to the partitioned feature $\mathbf{y}_i$, and the $\mathbf{y}_j$ from a previous step $j$ $(j<i)$ serves as a reference, enabling mutual adaptation. Finally, we develop a lightweight, fast-inference, high-performance neural analyzer $g_a(\cdot)$ and neural synthesizer $g_s(\cdot)$. The training methodology for this model will be presented in the following Section~\ref{subsec:training methodology}.

\begin{figure*}[htbp]
  \centering
  \includegraphics[width=0.99\textwidth]{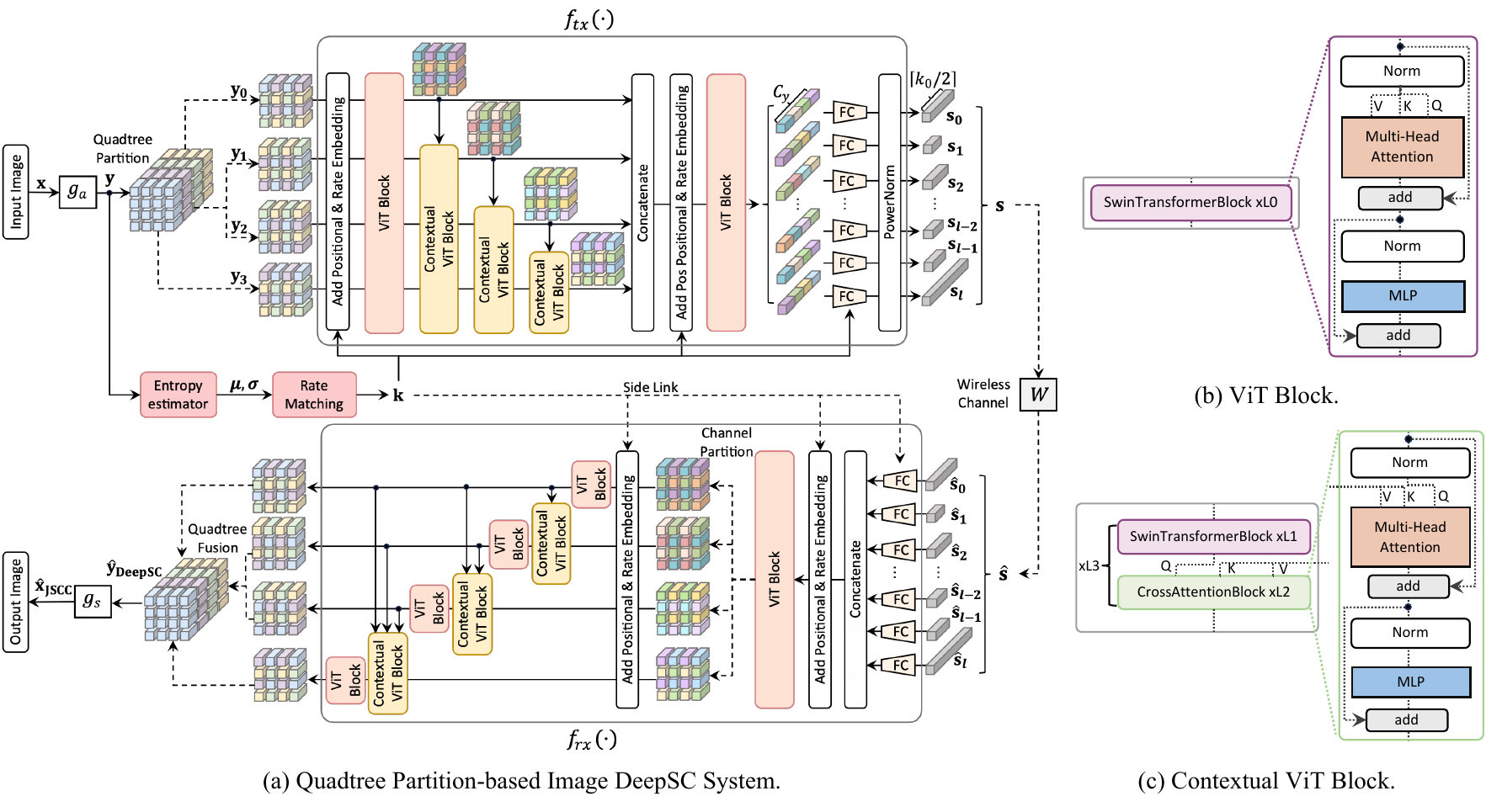}
  \caption{(a) Overview of proposed quadtree partition-based image DeepSC system (Quad-DeepSC). The side information $\mathbf{k}$ is derived from the estimated parameters $(\boldsymbol{\mu}, \boldsymbol{\sigma}^2)$ and transmitted via the side link to facilitate feature decoding, including symbol dimension mapping, positional and rate embedding selection, as detailed in Section~\ref{subsec:variable symbol length mapping}. (b) The network structure of the vision transformer (ViT) Block. (c) The network structure of contextual ViT Block. The module with the same name as in Fig.~\ref{overview of QSC} represents the identical component.}
  \label{overview of QJSCC}
\end{figure*}

\subsection{Model Architecture}
The detailed design of the proposed Quad-LIC is illustrated in Fig.~\ref{overview of QSC}(a). Notably, the Quad-LIC is implemented based on the ELIC framework~\cite{he2022elic}. In contrast to ELIC, the proposed approach discards its renowned entropy model and instead leverages a quadtree partition to construct a lightweight model, termed the quadtree partition-based space-channel context model (QP-SCCTX) that performs entropy estimation in merely four steps, thereby enabling fast coding speed. Despite its tendency to increase coding latency, the AttentionBlock in ELIC is retained to enhance the performance of the neural analyzer $g_a(\cdot)$ and neural synthesizer $g_s(\cdot)$. The corresponding structure is shown in Fig.~\ref{overview of QSC}(d). Additionally, a UNet, as illustrated in Fig.~\ref{overview of QSC}(b), is incorporated at the tail of $g_s(\cdot)$ to refine the reconstructed features. This enhancement improves the quality of source decoding and enhances robustness to channel impairments when the pre-trained Quad-LIC is embedded within the DeepSC system. For constructing both the QP-SCCTX and the UNet in Quad-LIC, we adopt techniques from DCVC-DC~\cite{dcvc-dc}, utilizing the DepthConvBlock illustrated in Fig.~\ref{overview of QSC}(c) as the basic block. In the the DepthConvBlock, except for the deep convolutional layers employing a $3\times3$ kernel, all regular convolutional layers utilize a $1\times1$ kernel to minimize computational complexity.

Building upon Quad-LIC, the proposed Quad-DeepSC introduces a feature encoder $f_{tx}(\cdot)$, a feature decoder $f_{rx}(\cdot)$, and a channel transfer function $W(\cdot)$. As shown in Fig.~\ref{overview of QJSCC}, we built the vision transformer (ViT) block as the basic block based on Swin-Transformer~\cite{liu2021swin}, owing to its demonstrated efficacy in generating noise-resilient channel symbol vectors $\mathbf{s}$~\cite{plit}. Furthermore, the relative position and rate configuration information can be effectively conveyed to the ViT Block by adding embedding to its inputs. To ensure consistency with entropy estimation, a cross-attention mechanism is introduced to construct the contextual ViT block~\cite{ntscc-plus}, which models contextual dependencies among DeepSC output codewords. Its structure is detailed in Fig.~\ref{overview of QJSCC}(c). The transformation of symbol dimensions is achieved using fully connected (FC) layers. The output symbols from the FC layers are power-normalized to satisfy the power constraint prior to wireless channel transmission.

\subsection{Quadtree Partition-Based Coding}
\label{subsec:quadtree partition-based coding}
\begin{figure}[!t]
  \centering
  \includegraphics[width=3.3in]{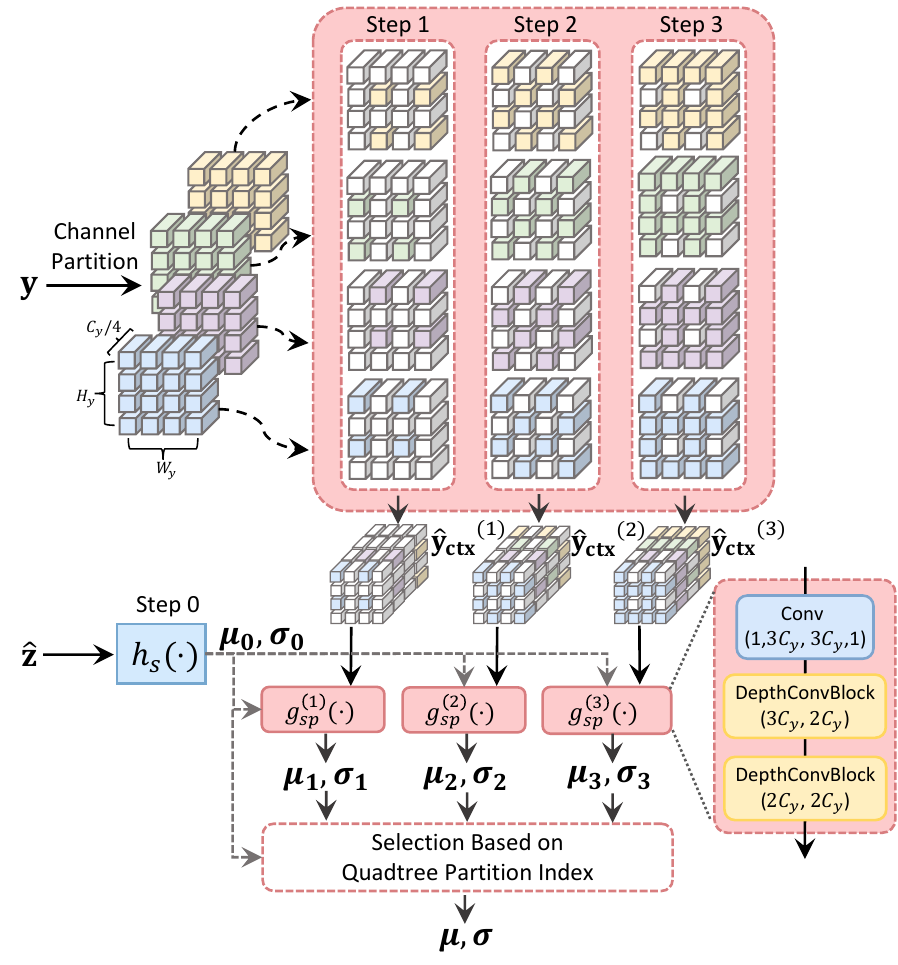}
  \caption{Diagram of the quadtree partition-based space-channel context model (QP-SCCTX).}
  \label{entropy model}
\end{figure}

We propose a quadtree-partitioned coding scheme that jointly designs an entropy estimator and feature coding. The entropy estimator comprises a hyper prior encoder $h_a(\cdot)$, a hyper prior decoder $h_s(\cdot)$ and QP-SCCTX $E_s(\cdot)$. The detailed architecture of proposed QP-SCCTX is illustrated in Fig.~\ref{entropy model}. Compared with the two-step checkerboard entropy model~\cite{he2021checkerboard}, QP-SCCTX provides finer granularity and more effectively exploits spatial contexts. Along the channel dimension, the semantic feature $\mathbf{y}$ is divided into four groups. Each element within a group is assigned an integer from 0 to 3 based on the spatial partition index defined by quadtree partition, which determines the sequential order of entropy estimation. Accordingly, at the $i_{>0}$th step, the context for estimating the current feature $\mathbf{y}^{(i)}$ is constructed from the previously estimated features $\mathbf{\hat{y}}_{ctx}^{(i)}$, thereby capturing both spatial and channel-wise dependencies:
\begin{equation}
  \label{eq:eq16}
    (\boldsymbol{\mu}_{i}, \boldsymbol{\sigma}^2_{i}) = g_{sp}^{(i)} \Big(\mathbf{\hat{y}}_{ctx}^{(i)}, (\boldsymbol{\mu}_{0}, \boldsymbol{\sigma}^2_{0}); \theta_{g_{sp}^{(i)}} \Big),
\end{equation}
where
\begin{equation}
  \label{eq:eq16-2}
  (\boldsymbol{\mu}_{0}, \boldsymbol{\sigma}^2_{0}) = h_s(\mathbf{\hat{z}}; \theta_{h_s}).
\end{equation}
Quantization is subsequently performed using the predicted entropy parameters:
\begin{equation}
  \label{eq:eq17}
  \mathbf{\hat{y}}^{(i)} = Q(\mathbf{y}^{(i)} - \boldsymbol{\mu}_{i}) + \boldsymbol{\mu}_{i}.
\end{equation}
Notably, spatial positions associated with index $i$ are non-overlapping within each group. Hence, at each step, one-quarter of the channels are estimated at each spatial location, and the previously estimated elements are concatenated to form the context for the next step. Across four steps, each element is progressively estimated with 0, 4, 4, and 8 neighboring references, respectively. In contrast, the checkerboard model~\cite{he2021checkerboard} uses 0 and 4 neighbors in the two steps. On average, QP-SCCTX leverages twice as many neighboring references, thereby enhancing inter-channel correlation modeling and enabling more accurate conditional probability estimation.

The semantic feature $\mathbf{y}$ contains redundancy along both the spatial and channel axes. To address this, we introduce quadtree partition-based feature coding scheme, as illustrated in Fig.~\ref{overview of QJSCC}(a). The estimation results of QP-SCCTX effectively guide feature coding to eliminate these redundancies while linking the output codewords of DeepSC to minimize CBR. $\mathbf{y}$ is partitioned into four groups: $[\mathbf{y_0}, \mathbf{y_1}, \mathbf{y_2}, \mathbf{y_3}]$ through quadtree partition. The probability distribution parameters of $\mathbf{y}_i$ is characterized by $(\boldsymbol{\mu}_{i}, \boldsymbol{\sigma}^2_{i})$, with its coding referencing the coding result of $\mathbf{y}_{<i}$ as context. This context corresponds to the reference $\mathbf{\hat{y}}^{(i)}_{ctx}$ for estimating $(\boldsymbol{\mu}_{i}, \boldsymbol{\sigma}^2_{i})$, allowing the feature coding to match the QP-SCCTX and exploit more granular and diverse contexts, thereby maximizing the utilization of correlations in both the spatial and channel dimensions. Furthermore, to provide $f_{tx}(\cdot)$ and $f_{rx}(\cdot)$ with position and probability distribution estimates related to $\mathbf{y}_i$, we compute the corresponding symbol length factor vector $\mathbf{k}_{i}$ for each element of $\mathbf{y}_i$, based on $(\boldsymbol{\mu}_{i}, \boldsymbol{\sigma}^2_{i})$. This $\mathbf{k}_{i}$ is then used to select the positional and rate embedding to be added to the inputs of $f_{tx}(\cdot)$ and $f_{rx}(\cdot)$.

\subsection{Variable Symbol Length Mapping}
\label{subsec:variable symbol length mapping}
Quad-LIC leverages arithmetic coding to inherently perform variable-length coding according to the estimated probability distribution. However, gradients are difficult to propagate through arithmetic codecs, and both CNN and Transformer architectures enforce fixed input-output channel dimensions, which limits the DeepSC systems from flexibly generating variable-length symbol sequences. A more computationally efficient approach is to utilize a set of FC layers to realize symbol mappings with variable lengths. Nonetheless, this introduces overhead in conveying to the receiver which FC layer was used for each symbol to restore dimensionality, thus incurring side information transmission costs. As shown in Fig.~\ref{overview of QSC}(a), the semantic feature $\mathbf{y}\in \mathbb{R}^{H_y \times W_y \times C_y}$, has a fixed channel dimension $C_y$, while the spatial dimensions $H_y = H/16$ and $W_y = W/16$ vary with input resolution $(H,W)$. Therefore, we choose the spatial unit $\mathbf{y}_{sp}^{(i,j)} \in \mathbb{R}^{C_y}$ as the fundamental entity for symbol mapping. On the one hand, the $C_y$ elements in each unit can interact during the mapping process. On the other hand, only $H_y \times W_y$ FC-layer indices need to be transmitted, reducing the overhead of side information.

In practice, we compute the length factor $k_{i,j}$ in (\ref{eq:kij}) for each corresponding symbol $\mathbf{s}_{i,j}$ by summing the entropy of the $C_y$ elements in $\mathbf{y}_{sp}^{(i,j)}$. To map $\mathbf{y}_{sp}^{(i,j)}$ from $C_y$ to $k_{i,j}$ dimensions, we use the shared FC layer $\text{fc}_{k_{i,j}}(\cdot)$ with weight shape $C_y \times C_y$, followed by a binary mask $\mathbf{m}_{k_{i,j}}$ to perform dimensionality reduction:
\begin{equation}
  \label{eq:eq18}
    \mathbf{r}_{i,j} = \text{fc}_{k_{i,j}}(\mathbf{y}_{sp}^{(i,j)}; \theta_{\text{fc}_{k_{i,j}}}) \odot \mathbf{m}_{k_{i,j}},
\end{equation}
where
\begin{equation}
  \label{eq:eq18-2}
  \mathbf{m}_{k_{i,j}} = [\underbrace{1,...,1}_{k_{i,j}}, 0,...,0 ] \in \mathbb{R}^{C_y}.
\end{equation}
The elements in $\mathbf{r}_{i,j}$ are paired to form the complex symbol $\mathbf{s}_{i,j}$. Similarly, $k_{i,j}$ is also used to select the positional and rate embedding. Consequently, $k_{i,j}$ constitutes a side information tensor $\mathbf{k} \in \mathbb{R}^{H_y \times W_y}$, whose transmission cost is given by:
\begin{equation}
  \label{eq:eq19}
    \text{CBR}_\text{side info} = \frac{b(\mathbf{k}) / C_\mathbf{k}}{H \times W \times 3} = \frac{H_y \times W_y \times \lceil \log_2(C_y) \rceil}{H \times W \times 3 \times C_\mathbf{k}},
\end{equation}
where $C_{\mathbf{k}}$ denotes the spectral efficiency of the side link to transmit $\mathbf{k}$. There is potential to optimize the side information cost in (\ref{eq:eq19}). Notably, $\mathbf{k}$ correlates with the semantic content and can be represented as a rate map on the $H_y \times W_y$ grid. By encoding this map using lossless portable network graphics (PNG), we can further reduce $b(\mathbf{k})$. Additionally, we restrict the range of $k_{i,j}$ using a quantization function $Q_k(\cdot)$ that maps each $k_{i,j}$ to its nearest value in a rate predefined set $[k_0, \dots, k_{n-1}]$, and transmit only the corresponding indices:
\begin{equation}
  \label{eq:eq20}
	\text{CBR}_\text{side info} = \frac{b\Big(\text{PNG}(Q_k(\mathbf{k}))\Big)}{H \times W \times 3 \times C_\mathbf{k}},
\end{equation}
where $Q_k(\cdot) \in [0, .., n-1]$ (e.g., $Q_k(k_i) = i$). This not only reduces $b\Big(\text{PNG}(Q_k(\mathbf{k}))\Big)$, but also limits the search space for FC layers and positional and rate embedding, thereby reducing computational complexity. The rate matching module in Fig.~\ref{overview of QJSCC}~(a) corresponds to the computation of $Q_k(\mathbf{k})$ based on the parameters estimated by the entropy estimator.

\begin{algorithm}[t]
  \caption{Training the Quad-DeepSC}
  \begin{algorithmic}[1]
    \State \textbf{Input:} 
    \Statex \hspace{1em} Training data $\mathbf{x}$, training steps for four stages $N_1$, $N_2$, $N_3$, and $N_4$, scaling hyperparameter $\eta$, lagrange multiplier $\lambda$, learning rate $l_r$.
    \State \textbf{Output:} 
    \Statex \hspace{1em} Parameters $(\theta_{g_a}^*, \theta_{g_s}^*, \theta_{h_a}^*, \theta_{h_s}^*, \theta_{g_{sp}}^*, \theta_{f_{tx}}^*, \theta_{f_{rx}}^*)$

    \State \textbf{Stage 1: Train the Learned Image Codec}
    \Statex Randomly initialize LIC parameters and fix feature encoder and decoder parameters, $(\theta_{f_{tx}}^*, \theta_{f_{rx}}^*)$.
    \Statex Approximate quantization with additive uniform noise.
    
    \For{$i \gets 1$ \textbf{to} $N_1$}
      \State Sample $\mathbf{x} \sim p_{\mathbf{x}}$
      \State Compute the loss function (\ref{eq:LIC_loss}): 
      \Statex \hspace{1em} $L_{LIC} = \lambda \cdot \mathcal{D}(\mathbf{x}, \mathbf{\hat{x}_{LIC}}) + R$
      \State Update the parameters: 
      \Statex \hspace{1em} $(\theta_{g_a}^*, \theta_{g_s}^*, \theta_{h_a}^*, \theta_{h_s}^*, \theta_{g_{sp}}^*)$
    \EndFor
    
    \State \noindent\rule{\linewidth}{0.4pt}  
    
    \State \textbf{Stage 2: Finetune the Learned Image Codec}
    \Statex Load parameters from Stage 1 and fix the feature encoder and decoder parameters.
    \Statex Apply actual quantization with gradient copying (STE).
    \Statex Lower the learning rate $l_r$.
    
    \For{$i \gets 1$ \textbf{to} $N_2$}
      \State Repeat steps 5 to 7 from Stage 1.
    \EndFor
    
    \State \noindent\rule{\linewidth}{0.4pt}  
    
    \State \textbf{Stage 3: Train the DeepSC System}
    \Statex Load LIC parameters from Stage 2.
    \Statex Approximate quantization with additive uniform noise.
    \Statex Apply quantization function $Q_k(\cdot)$ with a initial rate predefined set $[0, 1, ..., C_y]$.
    
    \For{$i \gets 1$ \textbf{to} $N_3$}
      \State Sample $\mathbf{x} \sim p_{\mathbf{x}}$
      \State Compute the loss function (\ref{eq:DeepSC_loss}): 
      \Statex \hspace{1em} $L = \lambda \cdot [\mathcal{D}(\mathbf{x}, \mathbf{\hat{x}_{LIC}} ) + \mathcal{D}(\mathbf{x}, \mathbf{\hat{x}_{DeepSC}} )] + K$
      \State Update the parameters: 
      \Statex \hspace{1em} $(\theta_{g_a}^*, \theta_{g_s}^*, \theta_{h_a}^*, \theta_{h_s}^*, \theta_{g_{sp}}^*, \theta_{f_{tx}}^*, \theta_{f_{rx}}^*)$
    \EndFor
    
    \State \noindent\rule{\linewidth}{0.4pt}  
    
    \State \textbf{Stage 4: Finetune the DeepSC System}
    \Statex Load parameters from Stage 3.
    \Statex Distill the rate predefined set $[0, 1, ..., C_y]$ to $[k_0, ..., k_{n-1}]$.
    \Statex Lower the learning rate $l_r$.
    
    \For{$i \gets 1$ \textbf{to} $N_4$}
      \State Repeat steps 17 to 19 from Stage 3.
    \EndFor
  \end{algorithmic}
  \label{alg1}
\end{algorithm}

\subsection{Training Methodology}
\label{subsec:training methodology}
The training procedure of Quad-DeepSC is presented in Algorithm~\ref{alg1}. Owing to the complexity of multiple distinct modules, direct end-to-end training may impede convergence. Therefore, the training for Quad-DeepSC is divided into four stages. Initially, the parameters $\theta_{f_{tx}}$ and $\theta_{f_{rx}}$ related to feature coding are frozen, and the Quad-LIC-related modules are optimized for source compression. The Quad-LIC training is divided into two stages. In the first stage, additive uniform noise is used to replace quantization and accelerate model convergence. In the second stage, actual quantization is applied with training conducted using the straight-through estimator (STE)~\cite{hu2023robust}. After these two stages, the Quad-LIC performance is evaluated. The performance of Quad-LIC is considered indicative of the overall DeepSC system performance. If Quad-LIC demonstrates slight size, fast inference and satisfactory performance, it implies that the designs of $g_s(\cdot)$, $g_a(\cdot)$, $E_a(\cdot)$, and $E_s(\cdot)$ are well-structured. Otherwise, additional ablation studies are required to eliminate redundant or ineffective components. Subsequently, we unfreeze $\theta_{f_{tx}}$ and $\theta_{f_{rx}}$, integrate the pre-trained Quad-LIC from the second stage into Quad-DeepSC, and conduct end-to-end training in two additional stages. Continuous quantization relaxation via additive uniform noise is preferable to discrete rounding for DeepSC: it preserves differentiability and yields smoother, more stable gradients over the stochastic channel, improving end-to-end optimization~\cite{pmlr-v139-guo21c, Pan_2021_CVPR}. Accordingly, we use the continuous surrogate during DeepSC training and reserve STE only for LIC validation. In the third stage, the quantization function $Q_k(\cdot) $ allows unrestricted selection of $k_{i,j}$ over the initial rate set $[0, \dots, C_y]$, with $C_y + 1$ FC layers and positional and rate embedding. In the fourth stage, the FC layers and embedding are distilled, restricting $k_{i,j}$ to a reduced rate set $[k_0, \dots, k_{n-1}]$. This reduces the side information cost, computational space complexity, and the search complexity. Furthermore, this distillation enables the selection of an optimal rate set tailored to each CBR range.

Notably, arithmetic codecs are only used during Quad-LIC evaluation. During training, the quantized semantic features $\mathbf{\hat{y}_{LIC}}$ are treated as arithmetic-decoding features. For training the Quad-DeepSC, additive uniform noise replaces quantization to alleviate the RDO problem. Additionally, channel impairments and the small batch size introduces gradient instability during training. To stabilize convergence, $\mathbf{\hat{y}_{LIC}}$ is directly passed to $g_s(\cdot)$, forming $\mathcal{D}(\mathbf{x}, \mathbf{\hat{x}_{LIC}})$ in the loss function~(\ref{eq:DeepSC_loss}), thereby anchoring entropy modeling and feature representation in Quad-LIC-related modules.

\section{Experimental Results}
\label{sec:experiments}
This section evaluates the performance of the proposed Quad-LIC and Quad-DeepSC through numerical experiments. We begin by detailing the experimental setup, followed by a presentation of the results using both visualizations and quantitative metrics.

\begin{figure*}[htbp]
  \centering
  \includegraphics[width=0.95\textwidth]{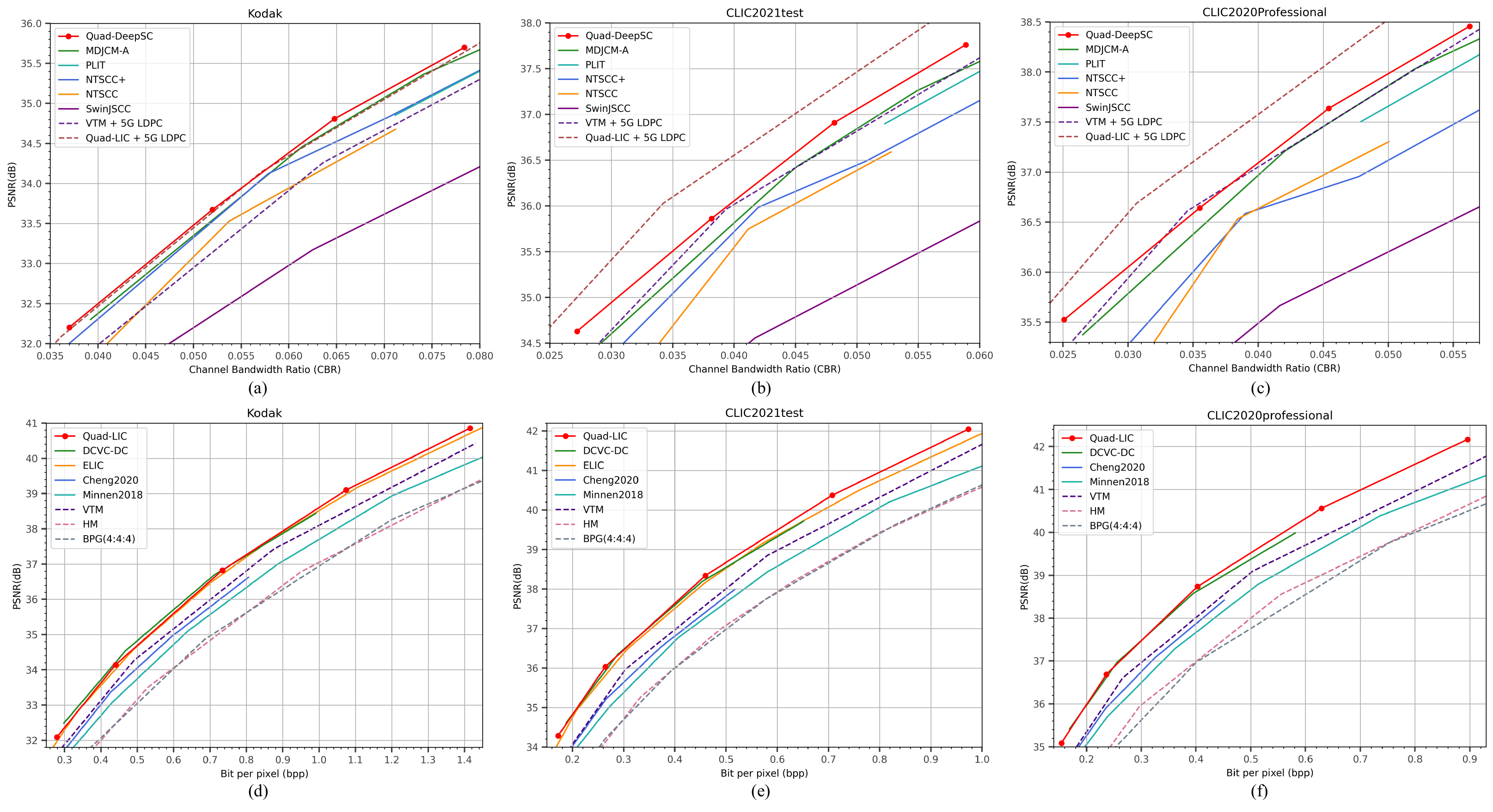}
  \caption{(a)-(c) PSNR performance of Quad-DeepSC for image transmission over the AWGN channel at SNR = 10 dB. (d)-(f) PSNR performance of Quad-LIC for image compression. All models are optimized to minimize MSE.}
  \label{PSNR}
\end{figure*}

\subsection{Experimental Settings}
\paragraph{Training Details} For training the Quad-LIC, we use the largest 8000 images picked from ImageNet~\cite{deng2009imagenet} dataset, after applying a noising-downsampling preprocessing~\cite{he2022elic, balle2018variational, he2021checkerboard}. For each architecture, models are trained with various $\lambda$ values for different quality presets. We use $\lambda=[6, 13, 35, 80, 150] \times 10^{-3}$ for each MSE-optimized model and $\lambda=[13, 35, 80, 150] \times 10^{-3}$ for each MS-SSIM-optimized model. The number of channels is set to $C_y=320$ and $C_z=192$ for all models. The Adam optimizer is employed with $\beta_1=0.9$, $\beta_2=0.999$, and images are randomly cropped to $256 \times 256$ patches. The model is initially trained with $\lambda=0.013$ to obtain an anchor model. The initial learning rate is set to $1e^{-4}$, with a batch size of 16, and uniform noise is sampled to estimate $\mathbf{\hat{y}}$ for the QP-SCCTX and synthesizer $g_s(\cdot)$. The anchor model is trained for 3800 epochs, followed by a learning rate decay to $1e^{-5}$ for an additional 200 epochs with $\mathbf{\hat{y}}=\text{STE}(\mathbf{y}-\boldsymbol{\mu})+\boldsymbol{\mu}$ for the QP-SCCTX and $g_s(\cdot)$. The anchor model is then fine-tuned using other $\lambda$ values, with a learning rate of $1e^{-5}$ and trained for 500 epochs. For training the Quad-DeepSC, since ImageNet is in JPEG format (a lossy version of the original images), we use the DIV2K~\cite{agustsson2017ntire} dataset, containing 800 high-resolution images. The trained Quad-LIC parameters are then used to initialize the Quad-DeepSC. The initial rate is set to $[0, 1, ..., C_y]$ and $\lambda=[40, 120, 240, 450] \times 10^{-3}$ for PSNR optimization, and $\lambda=[50, 240, 800, 2000] \times 10^{-3}$ for MS-SSIM optimization. The same optimizer is used, with a batch size of 8 and an initial learning rate of $1e^{-4}$. Each model is trained for 2000 epochs, followed by a learning rate decay to $1e^{-5}$ for another 1000 epochs.The rate set is then distilled to $[k_0, k_1, ..., k_{n-1}]$ and the model is fine-tuned with the same learning rate and batch size for 500 epochs. All experiments are conducted using PyTorch 2.4.1, Intel(R) Core(TM) i7-14700K @ 3.40GHz, and NVIDIA GeForce RTX 4090 GPU.

\begin{figure*}[htbp]
  \centering
  \includegraphics[width=0.95\textwidth]{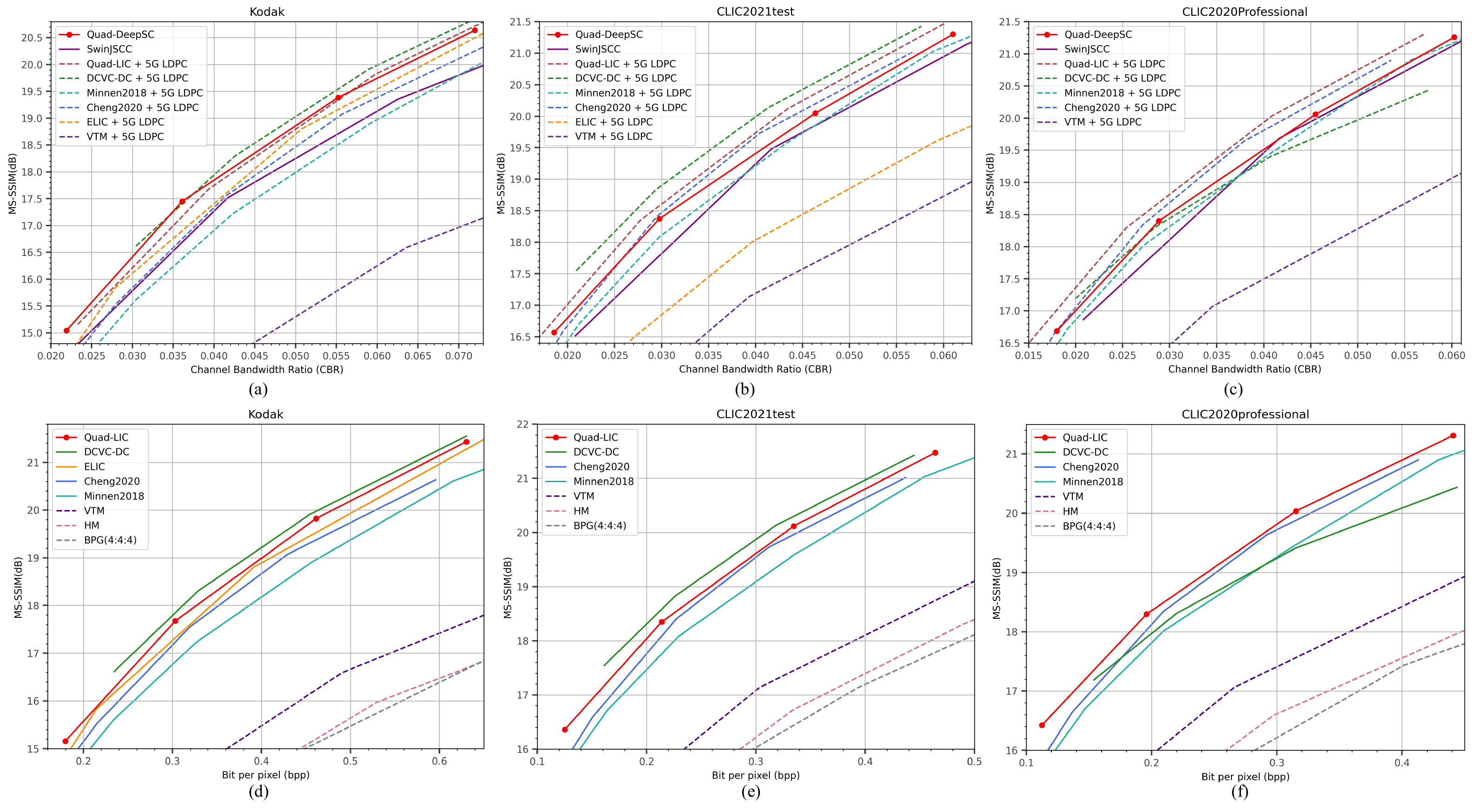}
  \caption{(a)-(c) MS-SSIM performance of Quad-DeepSC for image transmission over the AWGN channel at SNR = 10 dB. (d)-(f) MS-SSIM performance of Quad-LIC for image compression. All models are optimized to maximize MS-SSIM.}
  \label{MSSSIM}
\end{figure*}

\paragraph{Evaluation Settings} The performance of the proposed Quad-LIC and Quad-DeepSC models is evaluated based on PSNR and MS-SSIM. The test datasets include various resolutions. The Kodak~\cite{kodak1993} dataset, containing 24 images with resolutions of $512 \times 768$ or $768 \times 512$, the CLIC2020 professional testset~\cite{clic2020}, which includes 250 images with resolutions up to $2048 \times 1365$, and the CLIC2021 testset~\cite{clic2021}, which consists of 60 images with resolutions up to $2048 \times 1890$. All images are padded to multiples of 64 for evaluation. For benchmarks, we use the traditional image codecs such as BPG~\cite{bpg_bellard}, HEVC HM (still image codec)~\cite{hevc_hm}, and VVC VTM (still image codec)~\cite{vvc_vtm}, with the same test settings used in compressai~\cite{begaint2020compressai}. The emerging LICs include Minnen2018~\cite{minnen2018joint}, Cheng2020~\cite{cheng2020learned}, ELIC~\cite{he2022elic}, and DCVC-DC (still image codec)~\cite{dcvc-dc}. The end-to-end image DeepSC systems include NTSCC~\cite{ntscc}, NTSCC+~\cite{ntscc-plus}, MDJCM~\cite{mdjcm}, SwinJSCC~\cite{swinjscc}, and PLIT~\cite{plit}. We compare the traditional image codecs and LICs with 5G LDPC channel coding~\cite{richardson2018design}, with a length of 6144 for error correction and digital modulation, following the 3GPP standard. The optimal modulation order and channel code rate are chosen based on the MCS Table 5.1.3.1-1 for the PDSCH in~\cite{3gpp38214}. For each channel condition, we sweep the MCS index and adopt the setting that produces the lowest CBR. if decoding fails and the image cannot be reconstructed, we allow up to nine retransmissions. The metrics are computed on the first successful reconstruction. After testing all options, we report the results for the best configuration. Simulations are conducted using Sionna~\cite{sionna}. Source and channel coding schemes are concatenated with the “+" symbol. We losslessly compress $Q_k(\mathbf{k}) \in \mathbb{R}^{H_y \times W_y}$ with PNG. To ensure a fair comparison across DeepSC systems, we transmit the side information using the same scheme as the traditional baselines and report the resulting transmission cost.

\subsection{Results Analysis}
\paragraph{PSNR and MS-SSIM Performance} Fig.~\ref{PSNR} illustrates the PSNR performance of Quad-LIC and Quad-DeepSC. Specifically, Fig.~\ref{PSNR} (a)-(c) compare Quad-DeepSC with various transmission schemes under AWGN channel conditions at an SNR of 10 dB, while Fig.~\ref{PSNR} (d)-(f) depict the performance comparison between Quad-LIC and several existing image compression methods. The experimental results indicate that the proposed Quad-DeepSC significantly outperforms current DeepSC systems. Furthermore, the corresponding Quad-LIC achieves competitive results relative to existing LIC methods and significantly outperforms traditional image compression schemes.

To the best of our knowledge, this study is the first to propose an end-to-end transmission framework that surpasses the “VTM + 5G LDPC” benchmark on the CLIC2021 testset for 2K-resolution images, attaining a 2.53\% bandwidth reduction while maintaining comparable PSNR reconstruction fidelity. On the Kodak dataset, comprising medium-resolution images, Quad-DeepSC consistently surpasses the “VTM + 5G LDPC” benchmark across all considered CBR levels. Specifically, it achieves an 8.8\% reduction in bandwidth while preserving comparable reconstructed PSNR at the receiver. Quad-DeepSC also outperforms “Quad-LIC + 5G LDPC” scheme, highlighting the promise of DeepSC. However, on the CLIC2020 Professional testset, which contains many images above 2K resolution, all DeepSC methods underperform relative to “Quad-LIC + 5G LDPC” scheme. In that setting, Quad-DeepSC achieves a 0.8\% bandwidth reduction compared to the “VTM + 5G LDPC” scheme.

For better visualization, the MS-SSIM is converted into dB using the following formula: $-10\log_{10}(1 - \frac{\sum_{i=1}^N \text{MS-SSIM}_i}{N})$. Fig.~\ref{MSSSIM} displays the MS-SSIM performance of Quad-LIC and Quad-DeepSC, where Fig.~\ref{MSSSIM} (a)-(c) illustrate the comparison of various transmission schemes under the AWGN channel with 10 dB SNR, and Fig.~\ref{MSSSIM} (d)-(f) present the performance of different image compression techniques. The overall trend aligns with that observed in Fig.~\ref{PSNR}. Our Quad-LIC shows competitive performance compared to existing LIC schemes, significantly outperforming traditional image compression methods, while Quad-DeepSC exhibits competitive performance compared to existing “LICs + 5G LDPC" scheme. On the Kodak dataset, Quad-DeepSC achieves superior transmission performance over the “Quad-LIC + 5G LDPC" scheme, saving 3.3\% of bandwidth while maintaining the same MS-SSIM quality at the receiver.

\begin{table*}[htbp]
  \centering
  \caption{Rate-distortion (R-D) performance of various LICs on the CLIC2021 test dataset and computational complexity on the Kodak dataset, evaluated on an NVIDIA RTX 4090 GPU. Lower BD-Rate values indicate better R-D performance. The arithmetic codecs were run using two official GitHub repositories: one from~\cite{begaint2020compressai} and the other from~\cite{dcvc-dc}. The anchor is VTM. The best result is shown in $\textbf{bold}$.}
  \renewcommand{\arraystretch}{1.3}
  \begin{tabular}{>{\centering\arraybackslash}m{2.5cm}
                  |>{\centering\arraybackslash}m{3.2cm}
                  |>{\centering\arraybackslash}m{1.8cm}
                  |>{\centering\arraybackslash}m{1.8cm}
                  |>{\centering\arraybackslash}m{1.8cm}
                  |>{\centering\arraybackslash}m{1.8cm}
                  |>{\centering\arraybackslash}m{1.8cm}}
  \specialrule{.1em}{0pt}{0pt} 
  \textbf{LICs} & \textbf{Arithmetic Codec} & \textbf{BD-Rate (\%)} & \textbf{Enc time (ms)} & \textbf{Dec time (ms)} & \textbf{MACs (G)} & \textbf{Params (M)} \\
  \hline
  VTM & - & 0.0 & - & - & - & - \\
  \hline
  DCVC-DC~\cite{dcvc-dc} & \multirow{2}{*}{Cpp Rans Codec~\cite{dcvc-dc}} 
  & -11.8 & 15.7 & 22.0 & 217.34 & 31.00 \\
  Quad-LIC (proposed) &   & \textbf{-12.2} & \textbf{15.3} & \textbf{8.4} & 401.91 & 31.13 \\
  \hline
  Minnen2018~\cite{minnen2018joint} & \multirow{4}{*}{Compressai Rans Codec~\cite{begaint2020compressai}}& 10.2 & 1057.1 & 2343.2  & \textbf{176.85} & \textbf{25.5} \\
  Cheng2020~\cite{cheng2020learned} &  & 5.0 & 1077.2 & 2343.4 & 404.05 & 29.63 \\
  ELIC~\cite{he2022elic} &  & -9.3  & 95.2   & 59.1 & 327.88 & 33.79 \\
  Quad-LIC (proposed) &  & -12.2 & 40.5 & 36.8 & 401.91 & 31.13 \\
  \specialrule{.1em}{0pt}{0pt} 
  \end{tabular}
  \label{tab:sc_methods_fullpage}
\end{table*}

\begin{table*}[htbp]
  \centering
  \caption{Transmission performance of various DeepSC systems on the CLIC2021 test dataset and computational complexity on the Kodak dataset, evaluated on an NVIDIA RTX 4090 GPU. Lower BD-CBR values indicate better transmission performance. The anchor is “VTM + 5G LPDC" over the AWGN channel at SNR = 10 dB. The best result is shown in $\textbf{bold}$.}
  \renewcommand{\arraystretch}{1.3}
  \begin{tabular}{>{\centering\arraybackslash}m{3.5cm}
                  |>{\centering\arraybackslash}m{2.3cm}
                  |>{\centering\arraybackslash}m{2.3cm}
                  |>{\centering\arraybackslash}m{2.3cm}
                  |>{\centering\arraybackslash}m{2.3cm}
                  |>{\centering\arraybackslash}m{2.3cm}}
  \specialrule{.1em}{0pt}{0pt} 
  \textbf{DeepSC Systems} & \textbf{BD-CBR (\%)} & \textbf{Enc time (ms)} & \textbf{Dec time (ms)} & \textbf{MACs (G)} & \textbf{Params (M)} \\
  \hline
  VTM + 5G LPDC & 0.0 & - & - & - & - \\
  SwinJSCC~\cite{swinjscc}& 34.8 & 22.2 & \textbf{3.0} & \textbf{203.05} & \textbf{33.07} \\
  PLIT~\cite{plit}& 5.7 & 98.6 & 4.7 & 208.20 & 113.47 \\
  NTSCC~\cite{ntscc}& 17.9 & 20.3 & 3.3 & 262.80 & 31.36 \\
  NTSCC+~\cite{ntscc-plus}& 7.0 & 29.2 & 7.9 & 326.00 & 58.59 \\
  MDJCM-A~\cite{mdjcm}& 2.9 & 20.8 & 17.4 & 412.91 & 32.93 \\
  Quad-DeepSC (proposed) & \textbf{-2.5} & \textbf{12.1} & 16.2 & 422.64 & 50.58 \\
  \specialrule{.1em}{0pt}{0pt} 
  \end{tabular}
  \label{tab:qjscc_comparison}
\end{table*}

\paragraph{Computational Complexity} Table~\ref{tab:sc_methods_fullpage} summarizes the Bjøntegaard Delta Rate (BD-Rate. he average bitrate savings between two schemes over a range of qualities, measured at equivalent distortion. Lower is better)~\cite{bjontegaard2001calculation} results for PSNR, coding speed, and computational complexity of Quad-LIC, in comparison with existing LICs. Results for traditional image compression methods are excluded, as they are generally not optimized for GPU execution. For R-D evaluation, VTM is adopted as the baseline and tested on the 2K-resolution CLIC2021 dataset. Quad-LIC achieves an 12.2\% bitrate reduction without compromising PSNR quality, demonstrating its superior performance. The coding speed and computational complexity are evaluated on the Kodak dataset. When employing the arithmetic codec in Compressai (v1.1.5)~\cite{begaint2020compressai}, the encoding and decoding delays, along with computational space complexity during inference, are all lower than those of ELIC, achieving stronger performance owing to its lightweight entropy estimator and enhanced neural synthesizer with the added UNet module. Moreover, under the same arithmetic codec settings as DCVC-DC, the encoding delay of our Quad-LIC is reduced to 15.3 ms, and the decoding delay to 8.4 ms, significantly outperforming DCVC-DC in coding speed, while preserving better PSNR performance. This improvement stems from a more compact entropy estimator and a stronger feature transform network. Notably, by eliminating the computationally intensive RDO search process, Quad-LIC facilitates efficient GPU-based coding, offering substantial advantages for high-quality, real-time visual data transmission.

Table~\ref{tab:qjscc_comparison} presents the BD-CBR results for PSNR, coding speed, and computational complexity of Quad-DeepSC, compared to existing learned image transmission methods. For evaluating image transmission performance, the “VTM + 5G LDPC” scheme is adopted as the baseline and assessed on the 2K-resolution CLIC2021 test set. Quad-DeepSC achieves a noticeable reduction in bandwidth usage, while other DeepSC systems fail to yield any bandwidth savings. In terms of latency, Quad-DeepSC attains an encoding delay of 12.1 ms, making it the fastest among current DeepSC systems. The corresponding decoding delay is 16.2 ms, which is nearly symmetrical to the encoding latency. When transmission latency is considered, the convergence of these delays facilitates smooth and efficient wireless delivery of real-time visual data. This property constitute a distinctive advantage of the JSCC-based DeepSC paradigm, as traditional image codecs (e.g., VVC) require considerably longer encoding times, and channel codecs (e.g., 5G LDPC) introduce higher decoding delays.

\begin{figure*}[htbp]
\centering
\includegraphics[width=1.00\textwidth]{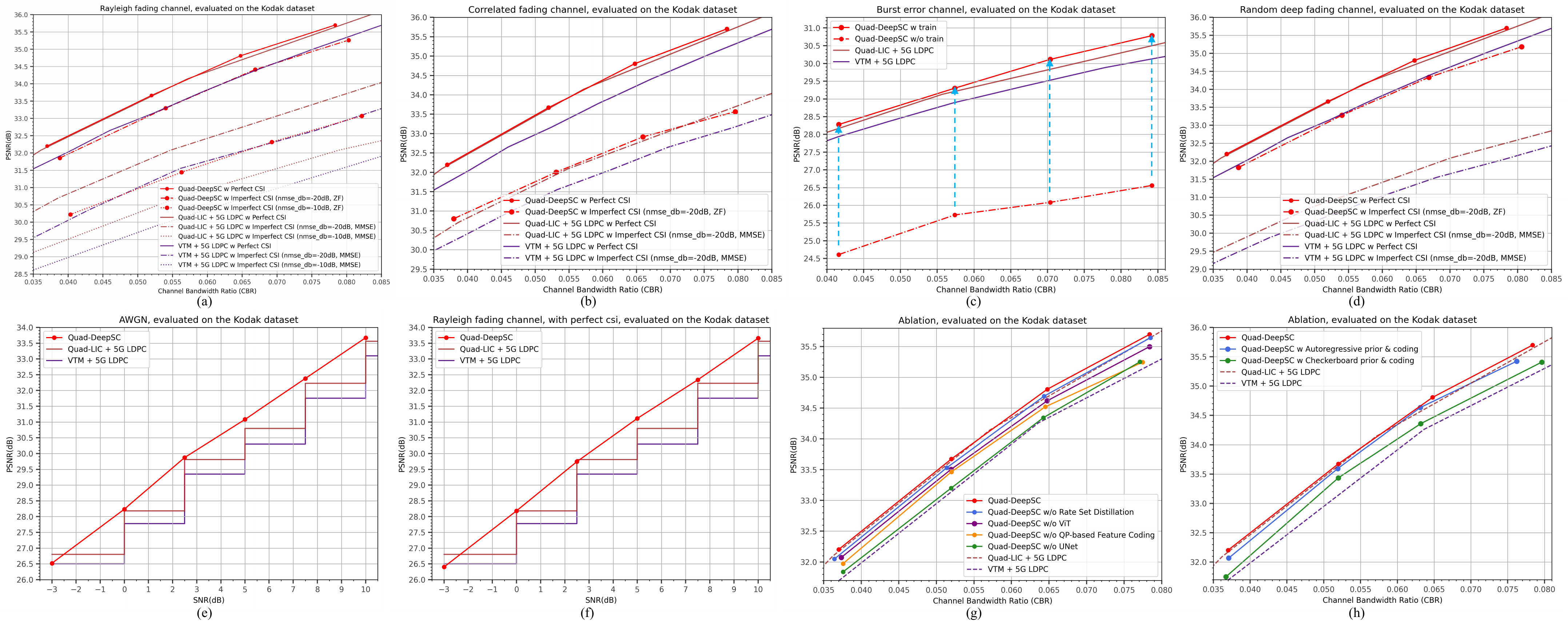}
\caption{PSNR of Quad-DeepSC under diverse channels at SNR $=10$ dB. (a)-(d) involve imperfect CSI with $\mathrm{NMSE}_{\mathrm{dB}}\in\{-20,-10\}$ dB. (e)-(f) use perfect CSI. (a) Rayleigh fading channel. (b) Correlated fading channel with coefficient $\rho=0.97$. (c) Burst-error channel with $p=0.05$ (good$\rightarrow$bad), $q=0.2$ (bad$\rightarrow$good), $\varepsilon=0.01$ (erase), and $\kappa=25$. (d) Random deep fading channel (LOS/NLOS blockages) with $p_{\text{block}}=0.02$, $L_m=50$, $K=10$ dB, $A_{\text{blk}}=20$ dB, and $\sigma_{\mathrm{sh}}=2$. (e)-(f) PSNR vs. SNR over AWGN and rayleigh fading channel at average CBR $=0.0515$. (g)-(h) Ablation studies on Quad-DeepSC over AWGN channel. All results are evaluated on the Kodak dataset.}
\label{Ablation}
\end{figure*}

\paragraph{Robustness Performance}
All results in Fig.~\ref{Ablation} are evaluated on the Kodak dataset with PSNR as the metric. Models are pre-trained on AWGN at 10 dB SNR and fine-tuned for 1,000 epochs with a learning rate of $1\times10^{-5}$ under the respective channel before evaluation. Fig.~\ref{Ablation} (a)–(d) report results at a fixed SNR of 10 dB. Fig.~\ref{Ablation} (e) and (f) sweep the SNR. When equalization is required, Quad-DeepSC uses ZF, while baselines use minimum MSE (MMSE).

Fig.~\ref{Ablation} (a) considers rayleigh fading. Under both perfect and imperfect CSI, Quad-DeepSC substantially outperforms the traditional schemes. With $\mathrm{NMSE}_{\mathrm{dB}}=-20$ dB, Quad-DeepSC matches the performance of “VTM + 5G LDPC” scheme under perfect CSI. With $\mathrm{NMSE}_{\mathrm{dB}}=-10$ dB, it remains comparable to “VTM + 5G LDPC” scheme at $\mathrm{NMSE}_{\mathrm{dB}}=-20$ dB, evidencing robustness to CSI degradation. Fig.~\ref{Ablation} (b) evaluates correlated fading with coefficient $\rho=0.97$, inducing moderately strong temporal correlation. Quad-DeepSC again exceeds the traditional schemes for both CSI settings and performs on par with “Quad-LIC + 5G LDPC” scheme. Fig.~\ref{Ablation} (c) studies a burst-error channel modeled by a two-state markov process with $p=0.05$ (good$\rightarrow$bad), $q=0.2$ (bad$\rightarrow$good), erase probability $\varepsilon=0.01$, and bad state noise inflation $\kappa=25$. Quad-DeepSC markedly surpasses traditional schemes. Moreover, targeted training further improves its performance under this impairment. Fig.~\ref{Ablation} (d) examines random deep fading via a LOS/NLOS process: LOS segments follow a rician model with $K=10$ dB. NLOS segments are rayleigh with additional amplitude attenuation $A_{\text{blk}}=20$ dB and log-normal shadowing standard deviation $\sigma_{\mathrm{sh}}=2$. Blockage entry probability is $p_{\text{block}}=0.02$ with mean blocked length $L_m=50$ symbols. Quad-DeepSC again outperforms the traditional schemes, with an even larger margin under imperfect CSI. Fig.~\ref{Ablation} (e) and (f) plot performance versus SNR from -3 to 10 dB over AWGN and rayleigh fading, respectively, at an average CBR of 0.0515 with perfect CSI. Across this range, Quad-DeepSC consistently exceeds the “VTM + 5G LDPC” benchmark. Overall, Quad-DeepSC adapts to diverse channel conditions and delivers robust end-to-end image transmission.

\subsection{Ablation Studies}
\label{subsec:ablation studies}

\begin{table*}[htbp]
  \centering
  \caption{Ablation studies on Quad-DeepSC, evaluated on the Kodak dataset with NVIDIA GeForce RTX 4090. The best and second-best results are shown in $\textbf{bold}$ and $\underline{\text{underlined}}$, respectively.}
  \renewcommand{\arraystretch}{1.15}
  \setlength{\tabcolsep}{3.8pt}
  \scriptsize
  \begin{tabular}{
    >{\centering\arraybackslash}m{3cm}|
    >{\centering\arraybackslash}m{1.0cm}|
    >{\centering\arraybackslash}m{1.0cm}|
    >{\centering\arraybackslash}m{1.6cm}|
    >{\centering\arraybackslash}m{1.0cm}|
    >{\centering\arraybackslash}m{1.0cm}|
    >{\centering\arraybackslash}m{1.0cm}|
    >{\centering\arraybackslash}m{1.0cm}|
    >{\centering\arraybackslash}m{1.0cm}|
    >{\centering\arraybackslash}m{1.0cm}|
    >{\centering\arraybackslash}m{1.0cm}|
    >{\centering\arraybackslash}m{1.0cm}
  }
    \toprule
    \textbf{} &
    \textbf{Enc time (ms)} &
    \textbf{Dec time (ms)} &
    \textbf{$g_a(\cdot)$+$p_{\hat{\mathbf y}}(\cdot)$ (ms)} &
    \textbf{$g_s(\cdot)$ (ms)} &
    \textbf{$f_{tx}(\cdot)$ (ms)} &
    \textbf{$f_{rx}(\cdot)$ (ms)} &
    \textbf{$Q_k(\mathbf{k})$ PNG enc (ms)} &
    \textbf{$Q_k(\mathbf{k})$ PNG dec (ms)} &
    \textbf{BD-CBR (\%)} &
    \textbf{MACs (G)} &
    \textbf{Params (M)} \\
    \midrule
    VTM + 5G LPDC
      & -- & -- & -- & -- & -- & -- & -- & -- & 0.0 & -- & -- \\
    \midrule
    Quad-DeepSC w/o UNet for $g_s(\cdot)$
      & 12.1 & 15.2 & \underline{3.8} & \textbf{1.4} & 8.3 & 13.8 & \textbf{0.218} & 0.231 & \phantom{-}0.8  & \textbf{382.74} & \underline{50.58} \\
    \midrule
    Quad-DeepSC w/o Rate set distillation
      & 13.0 & 16.6 & \underline{3.8} & \underline{2.4} & 9.2 & 14.2 & 0.240 & 0.241 & \underline{-7.0} & 422.64 & 68.15 \\
    \midrule
    Quad-DeepSC w/o Quadtree partition-based feature coding
      & \textbf{8.1} & \textbf{5.8} & \underline{3.8} & \underline{2.4} & \textbf{4.3} & \textbf{3.4} & 0.222 & \underline{0.222} & -3.1 & 422.47 & 54.23 \\
    \midrule
    Quad-DeepSC w/o ViT
      & \underline{8.3} & \underline{9.4} & \underline{3.8} & 4.1 & \underline{4.5} & \underline{5.3} & 0.226 & 0.236 & -4.9 & 421.71 & \textbf{50.52} \\
    \midrule
    Quad-DeepSC w/ Checkerboard prior \& coding
      & 10.7 & 8.1 & \textbf{3.6} & \underline{2.4} & 7.1 & 5.7 & 0.226 & \textbf{0.211} & -2.1 & 422.79 & 73.82 \\
    \midrule
    Quad-DeepSC w/ Autoregressive prior \& coding
      & 18.4 & 14.8 & 4.1 & \underline{2.4} & 14.3 & 12.4 & 0.243 & \underline{0.222} & -6.7 & \underline{420.78} & 65.88 \\
    \midrule
    Quad-DeepSC
      & 12.1 & 16.2 & \underline{3.8} & \underline{2.4} & 8.3 & 13.8 & \underline{0.221} & 0.233 & \textbf{-8.8} & 422.64 & \underline{50.58} \\
    \bottomrule
  \end{tabular}
  \label{tab:ablation}
\end{table*}

We first evaluate Quad-LIC in isolation for R-D performance, coding speed, and computational complexity. This serves as an ablation proxy for Quad-DeepSC, validating neural analyzer $g_a(\cdot)$, entropy estimator $p_{\mathbf{\hat{y}}}(\cdot)$, and neural synthesizer $g_s(\cdot)$. We then ablate Quad-DeepSC on Kodak dataset by removing or replacing key components, with emphasis on feature coding modules $f_{tx}(\cdot)$ and $f_{rx}(\cdot)$. Fig.~\ref{Ablation} (g) tests removal of rate set distillation (RSD), the UNet in $g_s(\cdot)$, ViT blocks in $f_{tx}(\cdot)$ and $f_{rx}(\cdot)$, and quadtree partition–based feature coding in $f_{tx}(\cdot)$ and $f_{rx}(\cdot)$. Removal is denoted “w/o” (e.g., “w/o UNet”). Fig.~\ref{Ablation} (h) compares partition strategies for entropy modeling and feature coding. Keeping the backbone fixed, we replace quadtree partitioning with checkerboard or channel-wise autoregressive alternatives in $p_{\hat{\mathbf y}}(\cdot)$, $f_{tx}(\cdot)$ and $f_{rx}(\cdot)$. Table~\ref{tab:ablation} summarizes the corresponding quantitative results.

Experimental results show that RSD improves performance, reduces memory consumption, and slightly accelerates the encoding process. Gains arise from a smaller search space for side information $\mathbf{k}$ and removal of redundant rate embeddings and FC layers. Removing the UNet in $g_s(\cdot)$ degrades reconstruction and robustness to channel impairments while reducing floating-point operations. We benchmark quadtree partition–based coding against checkerboard and autoregressive variants. Following~\cite{he2021checkerboard}, checkerboard variant splits feature into two groups. The checkerboard spatial indices ($i_0,i_1$) alternate every $C_y/4$ along channels (group 1: $[i_0,i_1,i_0,i_1]$, group 2: $[i_1,i_0,i_1,i_0]$). Autoregressive variant splits features into five groups as in~\cite{he2022elic} along channel dimension, with $C_y=320$ partitioned into $[16,16,32,64,192]$. Each partition scheme uses matching $p_{\hat{\mathbf y}}(\cdot)$, $f_{tx}(\cdot)$ and $f_{rx}(\cdot)$. From Fig.~\ref{Ablation} (h) and Table~\ref{tab:ablation}, quadtree partition–based coding delivers the largest gains and the lowest overall complexity. Autoregressive variant is the next best. Checkerboard variant performs weakest. For average speed, checkerboard variant is fastest and autoregressive variant slowest. Within the quadtree partition–based coding, replacing its feature coder with a uniform 4-layer ViT blocks shows entropy-adapted feature coding improves performance at a cost in speed. Conversely, swapping ViT blocks in $f_{tx}(\cdot)$ and $f_{rx}(\cdot)$ for DepthConvBlocks (all-CNN architecture) yields only a modest performance drop, indicating quadtree partition–based coding contributes more than ViT blocks. In addition, Table~\ref{tab:ablation} shows that the coding latency of the side information $\mathbf{k}$ is on the order of 0.1~ms, which is negligible compared with other modules. The quantized index $Q_k(\mathbf{k})$ is generated immediately after the entropy estimation of $p_{\mathbf{\hat{y}}}(\cdot)$. It can be transmitted concurrently with the operation of $f_{tx}(\cdot)$, thereby introducing no additional latency to the overall DeepSC system.

\section{Conclusion}
\label{sec:conclusion}
This paper proposed a deep learning based semantic communication system for images, inspired by quadtree partition as a feature segmentation method, named Quad-DeepSC. It is designed to balance rate-distortion performance and computational complexity, while being well suited for real-time wireless transmission of images with various resolutions. 
The core innovations of Quad-DeepSC are as follows. First, it employs quadtree partitioning for efficient feature coding and entropy modeling. Second, it introduces an optimized learned image codec (LIC), named quadtree partition-based LIC (Quad-LIC), which achieves excellent source compression performance with minimal coding latency. Third, it embeds Quad-LIC into the overall framework through an end-to-end training strategy, further enhancing system performance. 
Experimental results demonstrate that Quad-DeepSC surpasses the traditional “VTM + 5G LDPC” scheme in transmission efficiency, while maintaining low latency and strong robustness under diverse channel conditions. The Quad-LIC related components are CNN-based, whereas the feature coding module adopts a transformer structure. Considering that recent LIC increasingly utilize transformer or mamba architectures, future work will extend DeepSC toward these models to incorporate long-range dependency learning and dynamic state modeling, thereby improving coding efficiency.
In conclusion, Quad-DeepSC provides an ideal solution for efficient image transmission in future real-time communication scenarios. It has significant theoretical and practical value, effectively facilitating the integration of AI with communication systems and promoting the application of semantic communication.

\bibliographystyle{IEEEtran}
\bibliography{reference}

\end{document}